%% file: main.tex
\documentclass[journal]{new-aiaa}
\usepackage[utf8]{inputenc}
\usepackage{caption}
\usepackage{subcaption}
\usepackage{graphicx}
\usepackage{amsmath}
\usepackage[version=4]{mhchem}
\usepackage{siunitx}
\usepackage{longtable,tabularx}
\usepackage{placeins}
\usepackage{tikz}
\usepackage{soul}
\usepackage{xcolor}
\usepackage{booktabs}

\usetikzlibrary{shapes,arrows}

\title{A robust dynamic mode decomposition methodology for an airfoil undergoing transonic shock buffet}
\author{Andre Weiner\footnote{Post-doctoral researcher, flow modeling and control group, a.weiner@tu-braunschweig.de} and Richard Semaan.\footnote{Group leader, flow modeling and control group, r.semaan@tu-braunschweig.de}}
\affil{Technische Universit\"at Braunschweig, Institute of Fluid Mechanics, Braunschweig, 38108, Germany}

\tikzset{
base/.style = {rectangle, rounded corners, draw=black, fill=black!10,
               minimum width=13cm, minimum height=2.5cm, text centered},
optional/.style = {rectangle, rounded corners, draw=black, fill=black!10,
               minimum width=13cm, minimum height=2cm, text centered, dashed},
header/.style = {rectangle, draw=black, fill=white, minimum width=5cm,
                 minimum height=0.7cm, text centered}
}

\begin{document}

\maketitle

\begin{abstract}
Dynamic mode decomposition (DMD) is a data-driven technique widely used to analyze and model fluid problems, including transonic buffet flows.
Despite its strengths, DMD is known to suffer from sensitivities to the selected settings and the characteristics of the data used.
In this work, we closely examine the aforementioned sensitivities, identify possible pitfalls, and provide best practices for robustly performing DMD on a flow exhibiting transonic shock buffet.
Specifically, we assess several DMD variants and test their sensitivity to the POD rank truncation and the sampling rate. 
A critical enabler to our analysis is a new presentation of the DMD algorithm as a modular framework consisting of five distinct steps.
%The test results show a high sensitivity to the rank truncation.
The tests also highlight the existing dangers of aliasing, when the sampling rate is too low.
Finally, a list of practical recommendations and guidelines on how to accurately and robustly perform DMD on a transonic buffet flow is provided.
\end{abstract}

\section{Introduction}
\input{introduction}

\section{Dynamic mode decomposition}
\label{sec:dmd}
% \subsection{Singular value decomposition}
% \label{sec:svd}
\input{svd}

%\subsection{State vector}
%\label{sec:state}
%\input{state}

% \subsection{Dynamic mode decomposition}

\input{dmd}

\section{Test data}
\label{sec:datasets}

\input{dataset_oat15}

\section{Results}
\label{sec:results}
\input{results}

\section{Conclusions and recommendations}
\label{sec:summary}
\input{summary}

\section*{Appendix}
\label{sec:appendix}

\subsection*{Additional DMD variants}
\label{sec:dmd_add}

An alternative approach for de-noising applies a robust SVD on the data matrix decomposing it into low-rank $\mathbf{L}$ and sparse contributions $\mathbf{S}$ by solving the following optimization problem \cite{scherl2020}:
\begin{equation}
    \underset{\mathbf{L}+\mathbf{S} = \mathbf{M}}{\mathrm{argmin}}\left( \left|\left| \mathbf{L} ||_\ast + \lambda_0 || \mathbf{S} \right|\right|_1\right).
\end{equation}
The operator identification is subsequently performed on the low-rank contributions $\mathbf{L}$. 
While the TDMD approach requires only explicit computations, the robust SVD method involves a relatively costly iterative solution \cite{scherl2020}, which is why we exclude it from our current assessment study.

An alternative operator definition is the optimized DMD \cite{askham2018}, where the optimization problem is posed in terms of the scaled eigenvector matrix $\mathbf{\Phi}_b$ and the Vandermode matrix $\mathbf{V}_\mathbf{\omega}$
\begin{equation}
    \underset{\mathbf{\omega},\mathbf{\Phi}_b}{\mathrm{argmin}}\left|\left| \mathbf{X}-\mathbf{\Phi}_b\mathbf{V}_{\mathbf{\omega}} \right|\right|_F.
\end{equation}
The Vadermode matrix is based on the trace of the eigenvalue matrix, i.e., $\mathbf{\omega} = \mathrm{tr}\left(\mathbf{\Omega}\right)$; see appendix \ref{sec:continuous_operator} for more details. 
The reformulated optimization problem comes at the cost of a much more involved iterative solution procedure \cite{askham2018}. 
Even though the optimized DMD is usually more accurate \cite{askham2018,kaptanoglu2020}, the number of potentially important modes may increase, since the symmetry of complex-conjugate mode pairs is broken \cite{kaptanoglu2020}. 
Moreover, simple numerical experiments show that the resulting eigenvalues may be very different from that of the original dynamical system \cite{wu2021}. 
Due to the computational and numerical complexity of the optimized DMD, we do not consider this variant in our investigations.

An iterative approach to select meaningful modes is the sparsity-promoting DMD (spDMD) \cite{jovanovic2014}, which adds an $L_1$ norm to the optimization problem:
\begin{equation}
    \underset{\mathbf{b}}{\mathrm{argmin}}\left(\left|\left| \mathbf{X}-\mathbf{\Phi}\mathbf{D}_\mathbf{b}\mathbf{V}_{\mathbf{\omega}} \right|\right|_F+ \gamma_0\left|\left|\mathbf{b}\right|\right|_1\right)\,.
\end{equation}
The additional complexity of the spDMD requires an iterative solution procedure. 
We performed a priori tests with spDMD and found no advantage over optimal DMD
% a procedure with a non-sparse $\mathbf{b}$ 
when combined with the improved selection criteria, which is in agreement with the observations in \cite{kou2017}. 
Moreover, the spDMD introduces an additional hyperparameter $\gamma_0$ to control the sparsity, which requires tuning.

\subsection*{Details about the compressible inner product}
\label{sec:inner_product}

In this section, we provide some details on the choice of compressible inner products introduced by Rowley et al. \cite{rowley2004}.
The starting point is the total energy per unit mass, which is defined as the sum of internal energy $\hat{u}$ and kinetic energy $k$:
\begin{equation}
\label{eq:tot_energy}
    e = \hat{u} + k.
\end{equation}
The kinetic energy per unit mass in Cartesian coordinates is given by:
\begin{equation}
\label{eq:kin_energy}
    k = 0.5 \left( u_x^2 + u_y^2 + u_z^2 \right).
\end{equation}
Assuming ideal gas and isentropic flow, the internal energy may be expressed in terms of the local speed of sound $a$ \cite{rowley2004}:
\begin{equation}
\label{eq:int_energy}
    \hat{u} = \frac{a^2}{\gamma \left(\gamma - 1\right)},
\end{equation}
where $\gamma$ is the adiabatic index.
Substituting equations \eqref{eq:kin_energy} and \eqref{eq:int_energy} in \eqref{eq:tot_energy}, weighting by the density $\rho$, and integrating over the domain $\Omega$ yields:
\begin{equation}
\label{eq:total_energy}
    \int_{\Omega} \rho \left[
    \frac{a^2}{\gamma \left(\gamma - 1\right)} + 0.5 \left( u_x^2 + u_y^2 + u_z^2 \right)
    \right] \mathrm{d}v\approx \mathrm{const.}
\end{equation}
The integral is only approximately constant because the flow is not perfectly isentropic in the case of transonic shock buffet and there might be some netflux of energy across the boundary of $\Omega$.

\subsection*{Time-continuous linear operator}
\label{sec:continuous_operator}

In the time-continuous domain, the problem of advancing the state vector in time is posed as:
\begin{equation}
\label{eq:dmd_ode}
    \frac{\mathrm{d}\mathbf{x}}{\mathrm{d}t} = \mathbf{\mathcal{A}x},
\end{equation}
where $\mathbf{\mathcal{A}}$ is the continuous counterpart of $\mathbf{A}$. 
Both operators are related as:
\begin{equation}
    \mathbf{A} = \exp{\left(\mathbf{\mathcal{A}}\Delta t\right)},
\end{equation}
and similarly, the relation between the eigenvalues is:
\begin{equation}
    \mathbf{\Lambda} = \exp{\left(\mathbf{\Omega}\Delta t\right)},
\end{equation}
where the diagonal of $\mathbf{\Omega}$ contains the eigenvalues of $\mathbf{\mathcal{A}}$.
Equation \eqref{eq:dmd_ode} has an analytical solution \cite{kutz2014}:
\begin{equation}
    \mathbf{x}(t) = \mathbf{\Phi}\mathrm{exp}\left(\mathbf{\Omega} t\right) \mathbf{b}.
\end{equation}
Noticing that the eigenvalues $\omega_i$ in $\mathbf{\Omega}$ are usually complex numbers, the exponential term may be split up and reformulated using Euler's formula:
\begin{equation}
    \mathrm{exp}\left(\omega_i t\right) = \mathrm{exp}\left(\Re(\omega_i) t\right)\left[\mathrm{cos}\left(\Im(\omega_i)t\right) +j\mathrm{sin}\left(\Im(\omega_i)t\right)\right].
\end{equation}
Therefore, the $i$th eigenvalue's real part $\Re(\omega_i)$ is interpreted as growth rate of mode $i$, and the imaginary part $\Im(\omega_i)$ yields the mode's characteristic frequency $2\pi f_i = \Im(\omega_i)$.

\subsection*{UDMD buffet mode}
\label{sec:dmd_vs_udmd}

\begin{figure}[t!]
     \centering
     \includegraphics[width=0.8\textwidth]{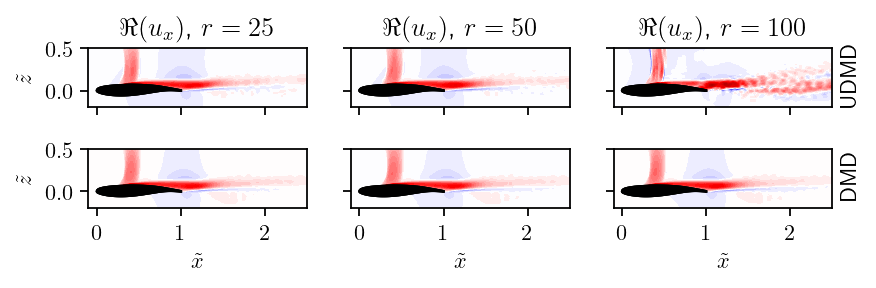}
     \caption{UDMD and DMD buffet modes for varying rank truncation.}
     \label{fig:udmd_modes_rank}
\end{figure}

Figure \ref{fig:udmd_modes_rank} shows the buffet modes resulting from UDMD for three different SVD ranks.
While the eigenvalue corresponding to the buffet mode remains nearly identical, the mode undergoes significant changes.
There is no obvious tendency for these changes regarding the selected rank, i.e. the effect is not caused by insufficient or exaggerated rank truncation. Presumably, enforcing the magnitude of the eigenvalues to be unity, shifts the impact of noise and nonlinear dynamics to the eigenvectors and causes the observed abrupt changes.
In contrast, the buffet mode resulting from the standard DMD operator remains almost identical over a wide range of $r$.

%%%
%%% Acknowledgments
%%%
\section*{Acknowledgments}

The authors gratefully acknowledge financial support by the German Research Foundation (DFG) received within the research unit FOR 2895 ``Unsteady flow and interaction phenomena at high-speed stall conditions''.
We also acknowledge the discussions with Steve Brunton, Benjamin Herrmann, and Peter Baddoo.
%%%
%%% References
%%%
\bibliography{literature}

\end{document}

%% file: introduction.tex
% general 2D shock buffet overview
\lettrine{T}{he} transonic shock buffet is a flow phenomenon characterized by self-sustained oscillations caused by interactions between the shock wave and the boundary layer over an airfoil under transonic flow conditions.
A good understanding of the shock buffet enables prediction of the flight envelope and, possibly, a mitigation of the phenomenon by means of flow control.
Numerous experimental and numerical studies have been conducted to deepen our understanding of the buffet mechanism.
A comprehensive review of the transonic shock buffet over airfoils and wings is provided by Giannelis et al. \cite{giannelis2017}. 
Below, we briefly summarize some aspects that are relevant to this work, which focuses on the robust data-driven analysis of the shock buffet rather than the intricacies of the buffet phenomenon itself. 
The two-dimensional shock buffet over airfoils is dominated by periodic shock motion along the chord combined with boundary separation downstream of the shock. 
The dominant buffet frequency in terms of the normalized frequency $\tilde{f}=2\pi f c/U_\infty$ ($c$: chord length, $U_\infty$: freestream velocity) is typically in the range $0.3 \leq \tilde{f} \leq 0.6$, corresponding to Strouhal numbers $St = fc/U_\infty$ in the range $0.05 \leq St \leq 0.1$. 
In addition to the dominant shock motion, instabilities in the separated boundary layer and vortex shedding at the trailing edge generate acoustic waves that propagate upstream and downstream \cite{hoffmann2021,jacquin2009,szubert2015}. 
Jacquin et al. \cite{jacquin2009} report propagation speeds of $-0.27U_\infty$ on the lower side and $0.072U_\infty$ on the upper side (the negative sign indicates upstream propagation). 
Above the separated boundary layer on the suction side, Hoffmann et al. \cite{hoffmann2021} also observe a propagation speed of $-0.36U_\infty$ and associate these acoustic waves with vortex shedding at the trailing edge. 
The reported shedding frequency is $\tilde{f}\approx 4.4$ ($St\approx 0.7$). 
Szubert et al. \cite{szubert2015} describe similar observations in their numerical study. 
However, they find the vortex shedding mode at a significantly higher frequency of $\tilde{f}\approx 15.7$ ($St\approx 2.5$). 
While Crouch et al. \cite{crouch2009} attribute the main buffet motion to an unstable global mode in their stability analysis, a potential two-way interaction between the shock motion and acoustic waves is still an active research topic.

The present investigations focus on the two-dimensional shock buffet. However, our observations and recommendations are likely to be applicable to flows exhibiting a so-called three-dimensional (3D) buffet, too. 
While the onset of both (2D) airfoil and (3D) wing buffets is associated with shock-induced perturbations, the two types of buffets have their own distinct mechanisms \cite{poplingher2023}. 
The 2D buffet is driven entirely by shock-wave boundary layer interaction, resulting in large chordwise shock oscillations.
On the other hand, the 3D buffet is associated with shorter shock motions resulting from a perturbation in the vicinity of the lambda shock that propagates towards the wing tip.
In contrast to the 2D buffet, which is dominated by the shock oscillation frequency, the 3D buffet exhibits a broad band of frequencies in the range $0.2 \leq St \leq 0.5$.
% While Lee \cite{lee1990} suggests a feedback loop between shock and vortex shedding driven by upstream propagating acoustic waves, more recently, Crouch et al. \cite{crouch2009} promoted the idea of the buffet as a global flow instability.

A better understanding of the underlying flow physics requires accurate and robust analyses.
One of the most widely used spectral analysis and reduced-order modeling techniques is the dynamic mode decomposition (DMD).
DMD has been repeatedly applied to transonic buffet flows \cite{ohmichi2018,poplingher2019,das2020,das2022,masini2020,hoffmann2021,zhao2020}.
However, DMD is known to suffer from sensitivities to the chosen settings and the data characteristics.
In this work, we benchmark various DMD variants and identify a robust methodology that yields physical and repeatable results.

% general DMD overview
The DMD is a purely data-driven modal decomposition technique \cite{schmid2010,kutz2014}.
Unlike similar techniques, such as proper orthogonal decomposition (POD), DMD associates each spatial mode with exactly one frequency and growth rate.
Characterizing the flow dynamics in terms of dominant frequencies is often more informative than a ranking based on second-order flow statistics \cite{schmid2011}, especially when acoustic waves need to be identified.
% Acoustic waves carry relatively little energy, such that a very large number of POD modes is required to capture them in turbulent flows \cite{freund2002}.
Despite its strengths, the DMD core algorithm is sensitive to noise and transients in the data \cite{wu2021} and requires careful parameter tuning, e.g., for the number of POD modes or regularization parameters.
Here, we use the term \textit{noise} generically for components in the data that are undesirable, e.g., because they have little or no coherence.
In this context, turbulent fluctuations in scale-resolved simulations are also considered noise.
To improve the accuracy and robustness of DMD, a number of extensions to the core algorithm have been developed \cite{schmid2022}. 
In the following, we review several DMD implementations applied to transonic shock buffet flows.

% overview on DMD for buffet analysis
Differences between the DMD implementations may be grouped according to (i) the type of data used, (ii) the definition of the state vector, and (iii) the employed DMD algorithm and its settings.
% , and (iv) insights obtained from the analysis.
The type of data is either numerical or experimental, 
which affects the amount of information available.
While simulations provide pressure, velocity, and other field variables across the entire flow domain, experiments typically yield only a single surface (e.g., via instationary pressure sensitive paint, iPSP \cite{masini2020}) or field (e.g., via particle image velocimetry, PIV \cite{hoffmann2021}) variable.
Moreover, simulations typically provide higher temporal and spatial resolution, while experiments provide much longer time sequences.
Table \ref{tab:data_sources} illustrates the typical differences in selected studies between the type of data used for DMD of a flow experiencing transonic buffet.
The ratio of sampling frequency $f_s$ and buffet frequency $f_b$ yields the number of snapshots per buffet cycle,
whereas $N_\mathrm{cycle}$ is the total number of buffet cycles recorded.
Note that $N_\mathrm{cycle}$ may or may not be equal to the number of cycles used in the modal analysis. It is not uncommon for numerical studies to use only a subset of the available snapshots to reduce the memory requirements.

\begin{table}[htbp]
    \centering
    \caption{Summary of selected literature resources illustrating the differences between the types of data used for DMD of a flow experiencing transonic buffet. The ratio of sampling frequency $f_s$ and buffet frequency $f_b$ yields the number of snapshots per buffet cycle. $N_\mathrm{cycle}$ is the total number of buffet cycles  recorded, and $\alpha$ is the angle of attack. 
    The $f_s/f_b$ values listed below for the 2D buffet \cite{poplingher2019,hoffmann2021} are calculated directly from the reported values. For the references investigating 3D buffet \cite{ohmichi2018,masini2020}, where the frequency of the dynamics is broadband, we have used the provided sampling frequency to estimate the lower and upper bounds of $f_s/f_b$, assuming a relevant interval of $0.2 \leq St_b \leq 0.5$.}
    % Reynolds and Mach number are defined as $Re=cU_\infty/\nu$ and $Ma=U_\infty / a_\infty$, respectively.}}
    \begin{tabular}{c c c p{3.2cm} p{5.3cm}}
        authors & $f_s/f_b$ & $N_\mathrm{cycle}$ & method and geometry & flow conditions \\\toprule
        Ohmichi et al. \cite{ohmichi2018} & $10-20$ & $\geq 20$ & 3D ZDES,\newline NASA CRM half-body & $Re_\infty = 1.516\times 10^6$,\newline $Ma_\infty = 0.85$, $\alpha = 4.87^\circ$ \\\midrule
        Massini et al. \cite{masini2020} & $8-25$ & $\approx 700$ & iPSP,\newline RBC12 half-body & $Re_\infty = 3.42\ldots 3.71\times 10^6$,\newline $Ma_\infty = 0.7\ldots 0.84$, $\alpha = 0.9\ldots 5.8^\circ$ \\\midrule
        Poplingher et al. \cite{poplingher2019} & $10-1000$ & $\geq 6$ & 2D URANS,\newline RA16SC1 airfoil & $Re_\infty = 4.2\times 10^6$,\newline $Ma_\infty = 0.65\ldots0.82$, $\alpha = 2\ldots 4^\circ$ \\\midrule
        Hoffmann et al. \cite{hoffmann2021} & $\approx 23$ & $\approx 265$ & PIV,\newline DRA2303 airfoil & $Re_\infty = 1.9\times 10^6$,\newline $Ma_\infty = 0.73$, $\alpha = 3.5^\circ$\\\bottomrule
    \end{tabular}
    \label{tab:data_sources}
\end{table}

The state vector may consist of volume, slice, or surface data of one or multiple field variables.
Simulations provide a larger degree of freedom in terms of possible state vectors. 
However, more complex state vectors in combination with high-resolution meshes can quickly lead to intractable memory requirements.
Ohmichi et al. \cite{ohmichi2018} used density, velocity, and pressure in a volume about a swept wing to define the state.
The authors also tested reduced state vectors containing only velocity or only pressure, and reported a comparable dominant buffet frequency in all cases.
Other numerical works \cite{poplingher2019,das2020,das2022} analyzed only the pressure field.
To better approximate the integral operation of the inner product and reduce the mesh dependence of the DMD results, Ohmichi et al. \cite{ohmichi2018} weighted the state vector with the square root of the cell volume.
% As detailed later in this manuscript, this weighting also reduces the mesh dependency on the results.
% While Ohmichi et al. \cite{ohmichi2018} did weight their state vector with the cell volume, a similar 
Such weighting was not reported by other authors \cite{poplingher2019,das2020,das2022}.
%In principle, the weighting should be also applied to iPSP data, since the surface area associated with a single pixel typically varies.
%However, to the best of our knowledge, this effect is typically ignored.

The final grouping of DMD applied to transonic buffet flows relates to the type of algorithm employed and its settings.
The first setting common to all DMD variants is the number of POD modes on which to project the linear operator.
While Ohmichi et al. \cite{ohmichi2018} truncated the POD basis after 51 modes without a clear justification, Das et al. \cite{das2020} chose the number of POD modes based on the reconstruction error.
The latter criterion may seem more rigorous at first glance.
However, a closer look reveals that the rank selection problem is now simply shifted to satisfy a user-defined reconstruction error.
% but essentially, the problem of choosing an appropriate rank is shifted to choosing an appropriate reconstruction error.
Other authors \cite{poplingher2019,masini2020,hoffmann2021,das2022} seem to have used the full POD basis, which is predetermined by the number of available snapshots.
However, the larger the POD basis, the more difficult the mode selection becomes.
Moreover, truncation of the POD basis is also important to reduce noise in the data.
To filter out important modes, some authors \cite{ohmichi2018,hoffmann2021} resorted to sparsity-promoting DMD (spDMD) introduced by Jovanović et al. \cite{jovanovic2014} or to a variant of it \cite{ohmichi2017}.
Poplingher et al. \cite{poplingher2019} employed the integral criterion of Kou and Zhang \cite{kou2017}, while Massini et al. \cite{masini2020} weighted the amplitudes with the eigenvalues as suggested by Tu et al. \cite{tu2014}.
The disadvantages of the spDMD are the increased cost that stems from the iterative solution procedure and the additional parameter tuning of the sparsity control parameter.
It is worth mentioning the recent multi-resolution DMD (mrDMD) implementation by Das et al. \cite{das2022}, where DMD is applied sequentially to increasingly smaller time windows, from which low-frequency dynamics are gradually removed \cite{kutz2016}.
% the authors perform an additional optimization step on the mode amplitudes according to Jovanović et al. \cite{jovanovic2014} and apply the DMD to smaller and smaller time windows, from which low-frequency dynamics were removed beforehand \cite{kutz2016} (multi-resolution DMD or mrDMD).
% The mrDMD breaks the pattern of finding only structures associated with the buffet frequency and its harmonics.
However, the extra effort provided little additional insight.

A common objective among most studies applying DMD to transonic buffet flows is their main focus on identifying the dominant buffet frequency and its harmonics.
Although important, the buffet mode is certainly not the only relevant one in the transonic buffet flow.
As noted above, Hoffmann et al. \cite{hoffmann2021} computed dominant modes at frequencies considerably higher than that of the buffet, 
which the authors attribute to vortex shedding at the trailing edge.
The authors also detected upstream traveling waves with a frequency similar to that of the vortex shedding along a streamline above the airfoil.
However, the acoustic waves were detected using spatio-temporal correlation coefficients rather than DMD.
A direct interaction between vortex shedding, acoustic waves, and shock motion has therefore not been established by DMD, even though DMD is in principle a suitable tool for investigating such relationships.
A notable attempt to investigate the interaction between the buffet mode and modes associated with vortex dynamics in the shear layer and wake regions was made by Szubert et al. \cite{szubert2015} on the OAT15A using POD.
The vortex shedding mode they identified, which the authors refer to as the Kármán mode, was characterized by a spectral bump rather than an isolated frequency, indicating frequency modulation.
The peak of this bump was located at about 33 times the buffet frequency.
The authors also report an influence of higher order harmonics of the buffet mode on the vortex shedding modes.
This smearing of the spectral content is not surprising, since the POD is not designed to identify modes with unique frequencies, but rather those with correlated phenomena. % that have some degree of coherence.
% The contribution of a POD mode to the temporal evolution of the flow is usually not characterized by an isolated frequency, but rather by a rich spectrum.
% In other words, a POD mode may encode a variety of flow dynamics that have some correlation.

% article structure, purpose of this article
The plethora of methods and our own analysis leading up to this work \cite{weiner2022} demonstrate the sensitivity of the DMD to the state vector definition, the rank truncation, the sampling rate, and the DMD variant employed.
This sensitivity makes the physical interpretation of the results and the quantitative comparison between studies extremely challenging.
In this work, we closely examine the aforementioned sensitivities, identify possible pitfalls, and provide best practices for performing DMD on flows exhibiting transonic shock buffet.
For the sake of reproducibility, we publish the complete work including simulation cases, datasets, processing pipelines, and visualizations in a dedicated repository\footnote{\url{https://github.com/FlowModelingControl/naca0012_shock_buffet}}.

The remainder of this article is structured as follows.
In section \ref{sec:dmd}, we recast DMD as a modular framework consisting of five steps with various options.
In section \ref{sec:datasets}, the simulation setup and data acquisition are briefly described.
In section \ref{sec:results}, we compare the outcome of various DMD variants, derive best practices, and apply the recommended procedure.
Our final assessment and recommendations are given in section \ref{sec:summary}.

%% file: svd.tex
Since its inception, many DMD variants have been proposed to improve its accuracy, robustness, efficiency, and applicability; see reference \cite{schmid2022} for a comprehensive review.
In this study, we are interested in identifying a DMD ``recipe'' that yields robust results
when subjected to noise and changes in the rank truncation.
Navigating through the plethora of existing DMD variants is a daunting task.
Therefore, we recast the dynamic mode decomposition as a modular five-step framework comprising (a) the definition of data matrices, (b) the optional denoising, (c) the operator identification, (d) the computation of the mode amplitudes, and (e) the selection of dynamically meaningful modes.
Figure \ref{fig:dmd_overview} illustrates this framework, 
wherein several DMD variants are denoted.
In principle, any option in one step can be combined with any option in another step.
However, not all combinations are sensible.
In this section, we focus on sensible and computationally tractable combinations.
Variants shown in gray in figure \ref{fig:dmd_overview} are not tested in this study. 
They are mentioned for completeness and are briefly described in the appendix \ref{sec:appendix}.
Note that we have structured this section according to the steps outlined in figure \ref{fig:dmd_overview}, 
which differs from the typical DMD textbook introduction.
This approach assumes a basic understanding of the core ideas of DMD. 
For an excellent introduction to DMD, we refer the reader to Kutz et al. \cite{kutz2014}.

% SVD definition
Before describing the new DMD framework in detail,
we will briefly review the singular value decomposition (SVD), which is an integral part of every DMD variant.
The SVD is a commonly used matrix factorization that yields the following decomposition
\begin{equation}
    \label{eq:SVD}
    \mathbf{X} = \mathbf{U}\mathbf{\Sigma}\mathbf{V}^\ast,
\end{equation}
where $\mathbf{U}$ and $\mathbf{V}$ are unitary matrices with orthonormal columns, and $\mathbf{\Sigma}$ is a diagonal matrix (considering the economy SVD).
Here, $^\ast$ denotes the complex conjugate transpose.
One of the most useful properties of SVD is that it provides an optimal low-rank approximation of a matrix $\mathbf{X}$, which can be  expressed mathematically as:
\begin{equation}
\label{eq:opt_rank}
    \underset{\mathbf{X}_r}{\mathrm{argmin}}\left|\left| \mathbf{X}-\mathbf{X}_r \right|\right|_F
    = \mathbf{U}_r\mathbf{\Sigma}_r\mathbf{V}^\ast_r
    \quad\textbf{subject to}\quad \mathrm{rank}(\mathbf{X}_r) = r,
\end{equation}
where $||\cdot ||_F$ is the Frobenius norm, and $r$ is the rank \cite{brunton2019}.
% $\mathbf{U}_r$ and $\mathbf{V}_r$ contain the first $r$ left and right singular vectors, respectively, and $\mathbf{\Sigma}_r$ is an $r\times r$ diagonal matrix holding the singular values.
% connection to POD and truncation
% When computing the DMD, the linear operator advancing the state in time is typically projected onto an orthonormal basis formed by the first $r$ POD modes, 
The rank $r$ is typically a user-defined parameter (a hyperparameter).
Choosing an appropriate value for $r$ is not always an easy task, especially for turbulent flows.
To facilitate rank truncation in the presence of noise, Gavish and Donoho \cite{gavish2014} proposed a simple set of equations to explicitly compute the optimal rank as:
\begin{equation}\label{eq:optRank}
    r_{\mathrm{opt}} = \underset{i}{\mathrm{argmin}}|\sigma_i - \kappa (\beta) \sigma_\mathrm{med}|,
\end{equation}
where $\sigma_\mathrm{med}$ is the median singular value, $\beta$ is the ratio of rows to columns of $\mathbf{X}$, and the function $\kappa (\beta )$ is given by
\begin{equation}
    \kappa (\beta ) = 0.56\beta^3 - 0.95\beta^2 + 1.82\beta + 1.43.
 \end{equation}
%  In practice, the index of the singular value closest to $\tau^\ast$ is then considered to be the optimal rank, denoted $r_{opt}$ hereafter.

%% file: dmd.tex
% Since its inception, many DMD variants have been proposed to improve the accuracy, robustness, efficiency, and applicability; see reference \cite{schmid2022} for a comprehensive review.
% Our focus is on using DMD as an analysis tool, which is why we care mostly about robustness.
% In particular, we are searching for DMD variants whose interpretability in terms of dominant modes and the corresponding spectrum changes as little as possible when subjected to noise and changes in the POD basis.
% To catalog the DMD variants considered in this study, we formulate the DMD as a five-step algorithm comprising the definition of data matrices, optional de-noising, operator identification, the computation of mode amplitudes, and the selection of dynamically meaningful modes.
% Figure \ref{fig:dmd_overview} provides an overview of multiple DMD variants according to the five-step categorization.
% In principle, each option in a step can be combined with each option in the following step.
% However, not all combinations are sensible.
% In the following sections, we review the options in each step and discuss potential advantages and disadvantages.

\begin{figure}
    \centering
    \input{overview_sketch}
    \caption{Schematic illustration of the modular DMD framework consisting of five distinct steps. Each step includes various choices, which lead to different DMD variants. We note that step (b) is optional and is therefore marked with a dashed line. Light gray text denotes variants that we did not test in this study.}
    \label{fig:dmd_overview}
\end{figure}

\subsection{Data matrices}
\input{data_matrices.tex}

\subsection{Pre-processing/de-noising}
\input{denoising.tex}

\subsection{Operator identification}
\label{sec:dmd_operator}
\input{operator.tex}

\subsection{Mode amplitudes}
\input{amplitudes.tex}

\subsection{Mode selection}
\input{mode_selection.tex}

\subsection{Abbreviations of DMD variants}

The various options in each step create a vast space of possible combinations.
To distinguish between the tested DMD variants, we use the abbreviations summarized in table \ref{tab:dmd_variants}. 
For example, ``opt. TDMD, int.'' refers to the DMD variant with total least-squares pre-processing, the standard operator identification, optimal mode coefficients, and mode ranking based on the integral mode selection criterion.
We emphasize that we restrict our tests to sensible and computationally-tractable combinations.

\begin{table}[t]
    \centering
        \caption{List of abbreviations of DMD options according to the framework illustrated in figure \ref{fig:dmd_overview}.}
    \begin{tabular}{c|c}
    Abbreviation & Explanation\\
    \hline
        DMD & standard operator definition \eqref{eq:least_squares} \\
        TDMD & total least-squares pre-processing \eqref{eq:tdmd} \\
        UDMD & unitary linear operator \eqref{eq:udmd} \\
        opt. & optimal mode amplitudes \eqref{eq:opt_dmd}\\
        amp. & mode selection based on $|b_i|$ \eqref{eq:standard_b}\\
        int. & mode selection based on the integral criterion \eqref{eq:int_criterion}
    \end{tabular}

    \label{tab:dmd_variants}
\end{table}

%% file: overview_sketch.tex
\begin{tikzpicture}

\node[base, minimum height=2.5cm] at (0, 0) {$
\mathbf{X} = \left[
\begin{array}{cccc}
|            & |            &     & | \\
\mathbf{x}_0 & \mathbf{x}_1 & ... & \mathbf{x}_{N-2} \\
|            & |            &     & | \\
\end{array}\right]\quad
\mathbf{Y} = \left[
\begin{array}{cccc}
|            & |            &     & | \\
\mathbf{x}_1 & \mathbf{x}_2 & ... & \mathbf{x}_{N-1} \\
|            & |            &     & | \\
\end{array}\right]
$};
\node[header] at (0, 1.25) {Data matrices};

\node[optional, minimum height=2.5cm] at (0, -3) {
%\begin{tabular}{c}
%$\mathbf{Z} = \left[
%\begin{array}{c}
%\mathbf{X} \\
%\mathbf{Y}
%\end{array}\right]$, $\mathbf{Z}=\mathbf{U}_{\mathbf{Z}r}\mathbf{\Sigma}_{\mathbf{Z}r} %\mathbf{V}_{\mathbf{Z}r}^\ast$, $\mathbf{P}_{\mathbf{Z}r} = %\mathbf{V}_{\mathbf{Z}r}\mathbf{V}_{\mathbf{Z}r}^\ast$\\[10pt]
%
%$\mathbf{X}\leftarrow \mathbf{XP}_{\mathbf{Z}r}$, $\mathbf{Y}\leftarrow \mathbf{YP}_{\mathbf{Z}r}$
%\end{tabular}
\begin{tabular}{cc}
\textbf{total least-squares} \cite{hemati2017} & \textcolor{gray}{\textbf{robust} \cite{scherl2020}}  \\[5pt]
$\underset{\mathbf{Y} + \mathbf{\Delta Y} = \mathbf{A}\left(\mathbf{X} + \mathbf{\Delta X}\right)}{\mathrm{argmin}}\left|\left| \left[\mathbf{\Delta X}, \mathbf{\Delta Y}\right]^T \right|\right|_F$ &
\textcolor{gray}{$\underset{\mathbf{L}+\mathbf{S} = \mathbf{M}}{\mathrm{argmin}}\left( \left|\left| \mathbf{L} ||_\ast + \lambda_0 || \mathbf{S} \right|\right|_1\right)$}
\end{tabular}
};
\node[header] at (0, -1.75) {Pre-processing/de-noising};

\node[base] at (0, -6) {
\begin{tabular}{ccc}
   \textbf{standard} \cite{kutz2014}  & \textbf{unitary} \cite{baddoo2021} & \textcolor{gray}{\textbf{optimized} \cite{askham2018}} \\[5pt]
    $\underset{\mathbf{A}}{\mathrm{argmin}}\left|\left| \mathbf{Y}-\mathbf{AX} \right|\right|_F$ &
    $\underset{\mathbf{A}^\ast\mathbf{A}=\mathbf{I}}{\mathrm{argmin}}\left|\left| \mathbf{Y}-\mathbf{AX} \right|\right|_F$ &
    \textcolor{gray}{$\underset{\mathbf{\omega},\mathbf{\Phi}_b}{\mathrm{argmin}}\left|\left| \mathbf{X}-\mathbf{\Phi}_b\mathbf{V}_{\mathbf{\omega}} \right|\right|_F$}
\end{tabular}
};
\node[header] at (0, -4.75) {Operator};

\node[base] at (0, -9) {
\begin{tabular}{cccc}
\textbf{standard} \cite{kutz2014} & \textbf{optimal} \cite{jovanovic2014} & \textcolor{gray}{\textbf{sparsity-promoting} \cite{jovanovic2014}} \\[5pt]
$\mathbf{b} = \mathbf{\Phi}^\dagger\mathbf{x}_0$ &
$\underset{\mathbf{b}}{\mathrm{argmin}} \left|\left|
\mathbf{X}-\mathbf{\Phi}\mathbf{D}_\mathbf{b}\mathbf{V}_{\mathbf{\omega}} \right|\right|_F$ &
\textcolor{gray}{$\underset{\mathbf{b}}{\mathrm{argmin}}\left(\left|\left| \mathbf{X}-\mathbf{\Phi}\mathbf{D}_\mathbf{b}\mathbf{V}_{\mathbf{\omega}} \right|\right|_F+ \gamma_0\left|\left|\mathbf{b}\right|\right|_1\right) $}&
\end{tabular}
};
\node[header] at (0, -7.75) {Mode amplitudes};

\node[base] at (0, -12) {
\begin{tabular}{ccc}
\textbf{amplitude} \cite{kutz2014} & \textbf{integral contr.} \cite{kou2017} &  \textcolor{gray}{\textbf{$\lambda$-weighted} \cite{tu2014}} \\[5pt]
$I_i = |b_i|$ & $I_i = \sum\limits_{j=0}^{N-1} |b_i\lambda_i^j|$ &  \textcolor{gray}{$I_i = |b_i \lambda_i^{N-1}|$}
\end{tabular}
};
\node[header] at (0, -10.75) {Mode selection};

\node[single arrow, draw=blue, very thick, fill=white, 
      minimum width=1cm, single arrow head extend=3pt,
      minimum height=1cm, rotate=-90] at (-5, -1.25) {};
\node[single arrow, draw=blue, very thick, fill=white, 
      minimum width=1cm, single arrow head extend=3pt,
      minimum height=1cm, rotate=-90] at (-5, -4.25) {};
\node[single arrow, draw=blue, very thick, fill=white, 
      minimum width=1cm, single arrow head extend=3pt,
      minimum height=1cm, rotate=-90] at (-5, -7.25) {};
\node[single arrow, draw=blue, very thick, fill=white, 
      minimum width=1cm, single arrow head extend=3pt,
      minimum height=1cm, rotate=-90] at (-5, -10.25) {};
      
\node[text=blue, minimum size=0.5cm,inner sep=0pt] at (5.5, 0.75) {\large $(a)$};
\node[text=blue, thick, minimum size=0.5cm,inner sep=0pt] at (5.5, -2.25) {\large $(b)$};
\node[text=blue, thick, minimum size=0.5cm,inner sep=0pt] at (5.5, -5.25) {\large $(c)$};
\node[text=blue, thick, minimum size=0.5cm,inner sep=0pt] at (5.5, -8.25) {\large $(d)$};
\node[text=blue, thick, minimum size=0.5cm,inner sep=0pt] at (5.5, -11.25) {\large $(e)$};

\end{tikzpicture}

%% file: data_matrices.tex
The first step in the modular DMD framework is the construction of two data matrices $\mathbf{X}$ and $\mathbf{Y}$ containing snapshots of the system state at fixed time intervals.
The snapshots in $\mathbf{Y}$ are shifted by one timestep with respect to $\mathbf{X}$.
% Starting from first principles, the temporal evolution of the flow variables in a spatial domain may be described by a set of conservation equations accompanied by suitable initial and boundary conditions.
Because SVD and DMD are data-driven methods, one could theoretically use any flow variable or a combination of multiple variables in the domain (or a subdomain) to define the state vector $\mathbf{x}$ \cite{schmid2022}.
However, it is typically advantageous to use state vectors that yield an interpretable and robust modal decomposition.
In compressible flows, the state vector could consist of density $\rho$, pressure $p$, velocity $\mathbf{u}=\left( u_x,u_y,u_z \right)^T$, or other derived field variables.
%  Following Rowley et al. \cite{rowley2004},
It is implicitly assumed that all flow states are functions of space.
The inner product of two states $\mathbf{x}_1$ and $\mathbf{x}_2$ is then defined as:
\begin{equation}
\label{eq:inner_product}
    \langle \mathbf{x}_1, \mathbf{x}_2 \rangle = \int_{\Omega} \mathbf{x}_1\cdot\mathbf{x}_2\> \mathrm{d}v,
\end{equation}
where $\Omega$ is the spatial domain under consideration.
The resulting norm is:
\begin{equation}
\label{eq:norm}
    ||\mathbf{x}||^2 = \int_{\Omega}  \mathbf{x}\cdot\mathbf{x}\> \mathrm{d}v.
\end{equation}
The norm plays an important role in the operator identification step.
For example, if we choose $\mathbf{x} = \left(u_x,u_y,u_z\right)^T$, the squared norm yields twice the kinetic energy of the flow in $\Omega$.
If the flow is incompressible, the kinetic energy is a preserved quantity and we could enforce this constraint in the DMD operator identification \cite{baddoo2021}.
% More details follow in section \ref{sec:dmd_operator}.

For compressible flows, where the state variables are coupled, Rowley et al. \cite{rowley2004} proposed a state vector that combines the velocity vector and the speed of sound as:
\begin{equation}
\label{eq:final_state}
    \mathbf{x} = \left(\sqrt{\frac{2}{(\gamma (\gamma - 1))}}\>a ,u_x, u_y, u_z\right)^T,
\end{equation}
where $\gamma$ is the adiabatic index and $a$ is the local speed of sound. 
The squared norm of definition \eqref{eq:final_state} yields twice the stagnation energy of the flow  \cite{rowley2004}, which is expected to be approximately conserved in the type of flows considered here. 
Further details on the derivation of the compressible inner product can be found in the Appendix \ref{sec:inner_product}.

% volume weighting
When dealing with snapshot data originating from a finite-volume solution, the state vector is finite-dimensional, and the inner product \eqref{eq:inner_product} is approximated by a sum over cell-centered values multiplied by the cell volume.
If the mesh is uniform, the volume-weighting can be neglected.
However, in the present study, the mesh around the airfoil is highly non-uniform, and neglecting the cell volume would introduce a strong mesh dependence in the DMD results.
The weighting is implemented in practice by multiplying each entry of the state vector by the square root of the corresponding cell volume.
The dot-product of two weighted state vectors then yields the desired approximation of the volume integral.

% time delays
 Besides the physically-motivated construction of the state vector, it is also possible to enrich the state with time delays by stacking multiple consecutive snapshots into extended column vectors \cite{brunton2016,clainche2017}.
 However, each additional snapshot increases the size of the final data matrices, which can be prohibitive for large datasets.
This problem can be significantly mitigated by a dimensionality reduction step prior to time delay embedding \cite{clainche2017}, yielding the so-called higher order DMD.
Time delay embedding is a very powerful enhancer that can be applied to any DMD variant. 
In this study, we do not consider any kind of time delay embedding.

%% file: denoising.tex
The second optional step deals with the often encountered small-scale broadband dynamics arising from turbulence.
% Small-scale turbulent structures typically have little to no coherence but are rather random. 
To reduce the sensitivity of the DMD analysis to the resolved turbulence, it may be sensible to apply a de-noising technique to the data before proceeding with the subsequent steps. 
Total least-squares DMD (TDMD) \cite{hemati2017} assumes that the data have some noise in $\mathbf{X}$ and $\mathbf{Y}$, denoted by $\Delta \mathbf{X}$ and $\Delta \mathbf{Y}$, which one attempts to minimize as,
\begin{equation}
\label{eq:tdmd}
    \underset{\mathbf{Y} + \mathbf{\Delta Y} = \mathbf{A}\left(\mathbf{X} + \mathbf{\Delta X}\right)}{\mathrm{argmin}}\left|\left| \left[\mathbf{\Delta X}, \mathbf{\Delta Y}\right]^T \right|\right|_F\,.
\end{equation}
While the linear operator $\mathbf{A}$ appears in the TDMD constraint, Hemati et al. \cite{hemati2017} present the solution of the total least-squares problem as a two-step procedure involving an initial subspace projection step followed by the operator identification. 
We adopt this view and consider the subspace projection as an optional pre-processing step before the operator identification. An alternative approach to de-noising is the robust DMD of Scherl et al. \cite{scherl2020}, which we outline in the appendix.

%% file: operator.tex
The third step identifies a linear operator $\mathbf{A}$ that maps the state $\mathbf{x}_n$ to $\mathbf{x}_{n+1}$, such that:
\begin{equation}
\label{eq:dmd_map}
    \mathbf{x}_{n+1} = \mathbf{A} \mathbf{x}_{n}.
\end{equation}
Since the exact linear operator will typically vary between snapshot pairs, $\mathbf{A}$ is defined in the least-squares sense as:
\begin{equation}
\label{eq:least_squares}
    \underset{\mathbf{A}}{\mathrm{argmin}}\left|\left| \mathbf{Y}-\mathbf{AX} \right|\right|_F.
\end{equation}
The solution of problem \eqref{eq:least_squares} without any further constraints on $\mathbf{A}$ yields the most common operator definition:
\begin{equation}
\label{eq:A_def}
    \mathbf{A} \coloneqq \mathbf{Y}\mathbf{X}^\dagger,
\end{equation}
where $\dagger$ denotes the Moore-Penrose inverse.

%The operator's eigen-decomposition, namely $\mathbf{A} = \mathbf{\Phi\Lambda\Phi}^{-1}$, has several useful properties. 
%For example, it can be used to make future predictions based on the relation:
%\begin{equation}
%    \mathbf{x}_n = \mathbf{\Phi\Lambda}^n \mathbf{b},
%\end{equation}
%where $\mathbf{\Phi}$ denotes the DMD modes and $\mathbf{b}=\mathbf{\Phi}^{-1}\mathbf{x}_0$ the mode amplitudes \cite{kutz2014}.
In fluid flow problems, the state vector is typically extremely large such that the eigendecomposition of $\mathbf{A}$ has to be approximated. 
The most widely used procedure invokes a low-rank approximation of $\mathbf{A}$ via an SVD of the first snapshot matrix, $\mathbf{X}\approx \mathbf{U}_r\mathbf{\Sigma}_r\mathbf{V}_r^\ast$ and the definition of a reduced operator
\begin{equation}
    \tilde{\mathbf{A}} = \mathbf{U}_r^\ast\mathbf{AU}_r = \mathbf{U}_r^\ast \mathbf{Y} \mathbf{V}_r\mathbf{\Sigma}_r^{-1}\,,
\end{equation}
which has the same non-zero eigenvalues as $\mathbf{A}$. 
The modes are then reconstructed according to the formulation proposed by Tu et al. \cite{tu2014}, yielding the so-called exact DMD modes:
\begin{equation}
    \mathbf{\Phi} = \mathbf{Y} \mathbf{V}_r \mathbf{\Sigma}_r^{-1}\mathbf{W},
\end{equation}
where the matrix $\mathbf{W}$ contains the eigenvectors of $\tilde{\mathbf{A}}$. Note that Tu et al. \cite{tu2014} scale the modes with the inverse of the eigenvalues $\mathbf{\lambda}$. As the authors point out, several other scalings are possible. 
We do not use any scaling and follow the more common definition of Kutz et al. \cite{kutz2014}.

This standard operator identification is the most widely used.
However, it still lacks accuracy and robustness for a number of applications.
Many variants have been proposed to address this shortcoming.

One variant we consider is from the physics-informed DMD family proposed by
Baddoo et al. \cite{baddoo2021}, where a number of physically-motivated constraints on the operator are introduced. 
One constraint we adopt in this study is the self-adjointness $\mathbf{A}^\ast \mathbf{A} = \mathbf{I}$, which assumes conservation of the state vector's norm.  
It is expressed mathematically as:
\begin{equation}
\label{eq:udmd}
  \underset{\mathbf{A}^\ast\mathbf{A}=\mathbf{I}}{\mathrm{argmin}}\left|\left| \mathbf{Y}-\mathbf{AX} \right|\right|_F.
\end{equation}
This constraint is only sensible if the state's norm is actually conserved (at least approximately).
We examine this assumption in section \ref{sec:weighting}.

Another variant is the optimized DMD \cite{askham2018}, 
which is described in the appendix.
This variant involves an optimization problem associated with a non-negligible computational overhead \cite{askham2018}. 
Although the optimized DMD is usually more accurate \cite{askham2018,kaptanoglu2020}, the number of potentially important modes may increase, since the symmetry of complex-conjugate mode pairs is broken \cite{kaptanoglu2020}. 
Furthermore, simple numerical experiments show that the resulting eigenvalues can be very different from those of the original dynamical system \cite{wu2021}. 
Due to the computational and numerical complexity of the optimized DMD, we do not consider this variant in our investigations.

%% file: amplitudes.tex
The fourth step concerns the computation of the mode amplitude vector $\mathbf{b}$.
The standard relation for computing the mode amplitudes results from the solution of the initial value problem \eqref{eq:dmd_ode} \cite{kutz2014}:
\begin{equation}
    \mathbf{b} = \mathbf{\Phi}^\dagger\mathbf{x}_0\ .
\end{equation}
However, since $\mathbf{A}$ is defined in the least-squares sense, there might be better options to compute the amplitudes, such as with optimal DMD \cite{jovanovic2014}, where
$\mathbf{b}$ is optimized for a given eigendecomposition: 
\begin{equation}
\label{eq:opt_dmd}
    \underset{\mathbf{b}}{\mathrm{argmin}} \left|\left|
\mathbf{X}-\mathbf{\Phi}\mathbf{D}_\mathbf{b}\mathbf{V}_{\mathbf{\omega}} \right|\right|_F\ .
\end{equation}

Another approach is the sparsity-promoting DMD (spDMD), which we briefly review in the appendix. 
The spDMD adds an $L_1$ norm to the optimization problem and thus more complexity and computational cost.
% \cite{jovanovic2014}, adds an $L_1$ norm to the optimization problem:
% \begin{equation}
%     \underset{\mathbf{b}}{\mathrm{argmin}}\left(\left|\left| \mathbf{X}-\mathbf{\Phi}\mathbf{D}_\mathbf{b}\mathbf{V}_{\mathbf{\omega}} \right|\right|_F+ \gamma_0\left|\left|\mathbf{b}\right|\right|_1\right)\,.
% \end{equation}
% The additional complexity of the spDMD requires an iterative solution procedure. 
We performed a priori tests with spDMD and found no advantage over optimal DMD
when combined with the improved selection criteria (presented in the next section), which is consistent with the observations in \cite{kou2017}. 
Moreover, the spDMD introduces an additional hyperparameter $\gamma_0$ to control the sparsity, which requires tuning.
For these reasons, we exclude spDMD from the current tests.

%% file: mode_selection.tex
The fifth and final step in the framework addresses the mode selection.
 Since the DMD modes are of unit length and the eigenvalues reflect the dynamic behavior (growing, decaying, periodicity), the magnitude of the amplitudes $b_i$:
 \begin{equation}\label{eq:standard_b}
     I_i = |b_i|\,.
 \end{equation}
can be used to sort the DMD modes.
 However, noise in the data often produces modes associated with large amplitudes and $|\lambda_i| < 1$, resulting in rapidly decaying modes that contribute very little to the full reconstruction.  
  To account for this damping, several authors \cite{tu2014,masini2020} rank the modes based on $\lambda_i^{N-1}$-weighted amplitudes as follows:
 \begin{equation}
      I_i = |b_i \lambda_i^{N-1}|\,.
 \end{equation}
 A more refined version of this idea was introduced by Kou and Zang \cite{kou2017},
 where the eigenvalue-weighted amplitudes are summed up over all $N$ time steps:
 \begin{equation}
 \label{eq:int_criterion}
     I_i = \sum\limits_{j=0}^{N-1} |b_i\lambda_i^j|\,.
 \end{equation}
Note that, compared to \cite{kou2017} we have omitted the multiplication by $\Delta t$, since it is not relevant for the ranking of the modes. We call this criterion the integral mode selection.

In contrast to the mode selection criteria presented so far, it is also possible to rank the modes based on their contribution to the reconstruction error, as by Ohmichi et al. \cite{ohmichi2018}. 
An exact solution of the associated combinatorial optimization problem is intractable. 
Even the greedy selection approach employed by Ohmichi et al. is still very expensive. 
Greedy mode selection means that at each iteration, the next most important mode is selected based on the prediction error in combination with the previously selected modes; see \cite{ohmichi2018} for more details. 
Therefore, selecting $k$ out of $r$ modes requires approximately $kr$ evaluations of the prediction error. 
Although we consider the greedy selection to be less practical, we nonetheless briefly compare it with the integral selection criterion in section \ref{sec:slice_analysis}.

%% file: dataset_oat15.tex
The simulation data we use to benchmark DMD variants were provided to us by a project partner within the research unit FOR 2895 \cite{lutz2022}. The simulation procedure and configuration is described in detail in Kleinert et al. \cite{kleinert2023}. 
In this section, we briefly describe the numerical setup and the data sampling procedure.

\begin{figure}[t]
    \centering
    \includegraphics[width=0.8\textwidth]{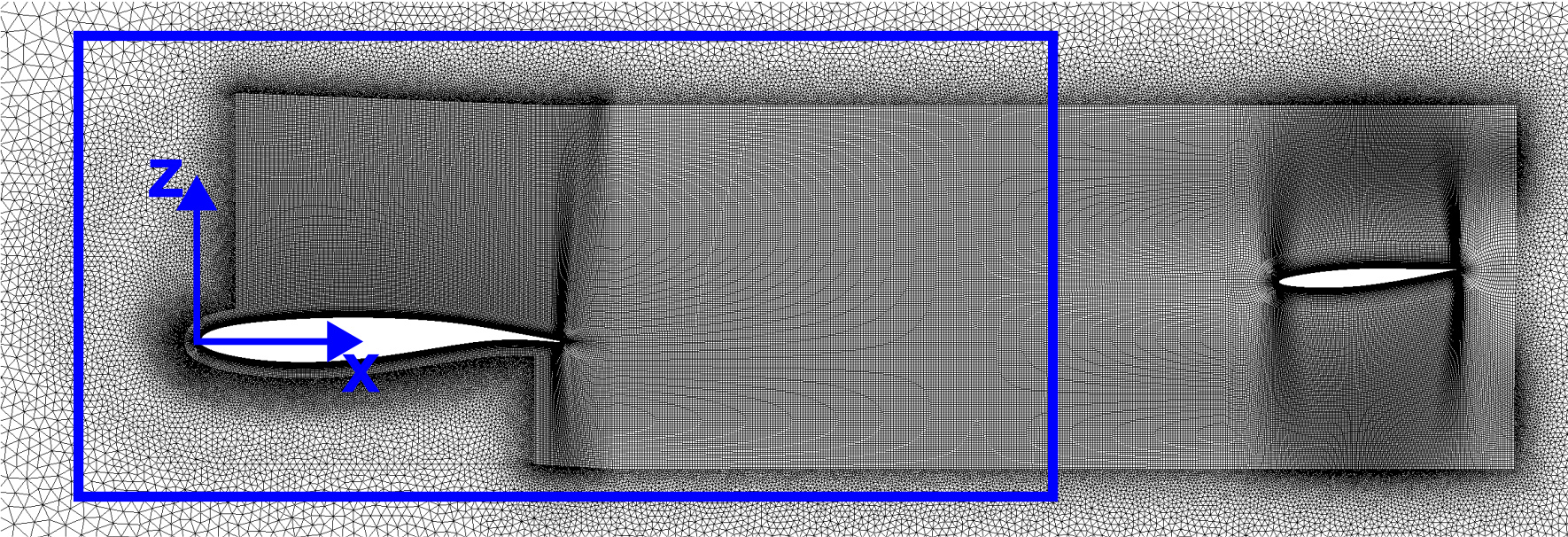}
    \caption{Hybrid mesh around the tandem configuration of OAT15A (front) and NACA 64A110 (rear). The blue frame indicates the reduced spatial domain used for the DMD analysis. Figure taken and modified from Kleinert et al. \cite{kleinert2023}.}
    \label{fig:tandem_mesh}
\end{figure}

Kleinert et al. \cite{kleinert2023} investigate a tandem configuration with an OAT15A airfoil in the front and a NACA64A110 airfoil in the rear. The distance between the trailing edge of the OAT15A and the leading edge of the NACA64A110 is $2c$. 
In the current study, we focus our attention on the OAT15A airfoil, which has a chord of $c=0.15\,$m. 
Therefore, we limit our spatial domain for the modal analysis to a plane defined by $-0.5 \leq \tilde{x} \leq 2.5$, $-0.5 \leq \tilde{z} \leq 1$, and $\tilde{y}=0$, where the normalized coordinates are $\tilde{x} = x/c$, $\tilde{y} = y/c$, and $\tilde{z} = z/c$.
This domain well-encompasses the shock and wake dynamics.
While it is clear that there is an effect of the leading airfoil's wake dynamics on the trailing airfoil, we made sure that there was no consequential interaction in the opposite direction, e.g., via acoustic waves.
To this end, we performed modal analyses of the OAT15A in isolation, the NACA64A110 in isolation, and the full tandem configuration.
Comparison of the resulting spectra revealed no trailing edge frequency content in the OAT15A spectrum.
The origin of the coordinate system is at the leading edge of the OAT15A airfoil. 
This airfoil configuration along with the reduced spatial domain is shown in figure \ref{fig:tandem_mesh}.

The mesh in the vicinity of the airfoils, the shock region, and the wake of the leading airfoil consists of hexahedral elements. The remainder of the hybrid mesh is composed of triangular prisms. The farfield boundary of this O-type mesh is located at a distance of approximately $50\,$m, which corresponds to $95$ times the length of the full tandem configuration. In the spanwise direction, $\tilde{y}$, the domain width is $0.49c$. The domain boundaries at $ \tilde{y}=0 $ and $ \tilde{y}=0.49 $ are periodic. 
The first cell layer around the OAT15A airfoil is constructed so that $y^+ < 0.4$ is guaranteed over the entire surface, where $y^+$ is the surface normal coordinate in wall units. Along the chord, the resolution is nearly constant with an element size of $0.004c$. The element size in the wake is about $0.07c$. The mesh in spanwise direction is generated by extrution of an initial 2D mesh. The 70 cell layers yield a spanwise resolution of $0.07c$. In total, the mesh consists of $\approx 16\times 10^6$ cells.

The numerical solution is obtained using the TAU flow solver \cite{dlr55519} with AZDES (automated zonal detached eddy simulation) turbulence modelling \cite{ehrle2020}. A SSG/LRR-$\omega$ Reynolds stress model serves as the turbulence model in the RANS and as subgrid-scale model in the LES regions. The boundary layer is tripped at $7\%$ chord of the OAT15A profile. The numerical timestep is $\Delta t = 4.4\times 10^{-6}\,$s, corresponding to 150 steps per convective time unit $\tilde{t}_c=c/U_\infty$.
The numerical method has been used and validated in previous studies of transonic shock buffets \cite{ehrle2020,ehrle2020b,Kleinert2023a}.

Table \ref{tab:flow_cond} summarizes the flow conditions and the data acquisition. 
Again, for the modal analysis, we use only the periodic boundary at $\tilde{y}=0$ within the region marked in figure \ref{fig:tandem_mesh}. This plane consists of $N_\mathrm{cells} \approx 1.5\times 10^5$ quadrilateral and triangular elements. All flow variables are saved at  $10\Delta t$ intervals over a period of two buffet cycles yielding a total of $N_\mathrm{snap} = 400$ snapshots. However, because we are performing a computationally intensive parametric study, we have sampled the data down to every second snapshot, corresponding to $f_s/f_b = 100$. Only for the samling rate sensitivity analysis in section \ref{sec:sampling_rate} is the ratio $f_s/f_b$ varied.

\begin{table}[t]
    \centering
    \begin{tabular}{c c c c c c}
       $Re_\infty$ & $Ma_\infty$ & $\alpha$ & $N_{cycle}$ & $N_{cells}$ & $N_{snap}$ \\\toprule
       $2\times 10^6$ & $0.72$ & $5^\circ$ & $2$ & $1.5\times 10^5$ & $400$
    \end{tabular}
    \caption{Summary of the flow conditions and data acquisition for the tandem configuration. $N_\mathrm{cells}$ is the number of cells in the reduced domain used for the DMD analysis; see the blue rectangle in figure \ref{fig:tandem_mesh}. $N_\mathrm{snap}$ is the number of available snapshots.}
    \label{tab:flow_cond}
\end{table}

Based on the lift coefficient $c_l$, Kleinert et al. \cite{kleinert2023} report a buffet frequency of $f_b = 118.5 Hz$, which corresponds to a normalized frequency of $\tilde{f}_b=0.46$ ($St=0.0745$) and is in the expected range of $0.3 \leq \tilde{f} \leq 0.6$.
The buffet frequency agrees well with the numerical study of an isolated OAT15A profile by Szubert et al. \cite{szubert2015}, who report $\tilde{f}_b = 0.47$ at $\alpha = 3.5^\circ$. 
The lift fluctuation amplitude $0.81 \leq c_l \leq 1.10$ also agrees well with the range of $0.68 \leq c_l \leq 1.08$ in \cite{szubert2015}, taking into account the difference in the angle of attack.
Compared to the experimentally determined buffet frequency by Jacquin et al. \cite{jacquin2009} ($\tilde{f}_b\approx 0.39$) under slightly different flow conditions ($Re_\infty = 3\times 10^6$, $Ma_\infty = 0.72$, $\alpha = 3.5^\circ$), the numerical prediction yields a $15\%$ higher frequency.
Besides the buffet frequency, Kleinert et al. \cite{kleinert2023} report dominant frequencies in the range $12.6 \leq \tilde{f} \leq 16.3$ ($2.0 \leq St \leq 2.6$) based on a spectral analysis of both POD mode coefficients and surface pressure. This is again consistent with the frequency peak of $\tilde{f}=15.7$ ($St=2.5$) reported in \cite{szubert2015}.

%% file: results.tex
% \subsection{Spectra at various probe locations}
% \label{sec:spectra_probes}

\input{probes.tex}

\subsection{State vector weighting}
\label{sec:weighting}
\input{weighting}

\subsection{Rank sensitivity}
\label{sec:rank_sensitivity}
\input{rank_sensitivity.tex}

\subsection{Sampling rate sensitivity}
\label{sec:sampling_rate}
\input{sampling_rate_sensitivity.tex}

%\subsection{Time window sensitivity}
%\label{sec:sampling_time}
%\input{sampling_length_sensitivity.tex}

\subsection{Application}
% \subsubsection{DMD on a slice of data}
\label{sec:slice_analysis}

\input{slice_analysis.tex}

% \subsubsection{DMD on surface and volume data}
% \label{sec:surface_analysis}
% \input{surface_volume_analysis.tex}

%% file: probes.tex
% In the previous section, we briefly reviewed various dynamic mode decomposition variants and their relations.
In this section, we test different DMD variants with different settings on the transonic buffet flow.
We assess each algorithm based on its robustness and its accuracy in identifying the flow dynamics. 
Robustness tests include rank sensitivity and sampling rate sensitivity.
Using the test results, we identify a robust recipe and apply it to a planar slice through the domain.
%Unless otherwise stated, the state vector in all tests is based on the stagnation energy.
We note that we have performed the same analyses on the publicly available geometry of the NACA 0012, but these results are omitted here for brevity.
The interested reader is referred to the accompanying repository.
%Using the test results, we identify a robust recipe, apply it to the flow surface and field variables, and draw some conclusions.
%The analysis is conducted on a planar slice through the domain employing the physically-motivated state vector that preserves the stagnation energy (equation \eqref{eq:final_state}).
All DMD analyses are conducted using the Python library flowTorch \cite{weiner2021}.

%% file: weighting.tex
\begin{figure}[t]
    \centering
    \includegraphics[width=0.8\textwidth]{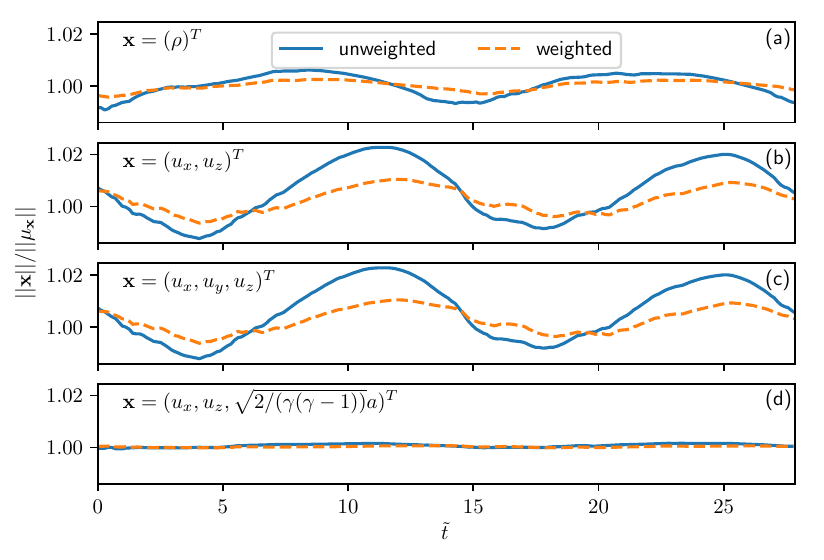}
    \caption{Time sequence of various state vector norms $||\mathbf{x}||$ divided by their corresponding temporal mean norm $||\mu_\mathbf{x}||$. The legends indicate the field variables comprising the state vector. The keyword \textit{weighted} indicates a weighting with the square root of the face area.\label{fig:state_norms}}
\end{figure}

Figure \ref{fig:state_norms} presents the temporal evolution of various state vector norms divided by the norm of their corresponding temporal mean state.
The state vector norms for field variables are presented with and without cell volume/area weighting (volume and area weighting are equivalent in the present study as the 3D mesh is created by extrusion).
To assess the applicability of the unitary operator constraint \eqref{eq:udmd}, we analyze the temporal evolution of typical state vector norms.
%Figure \ref{fig:state_norms} shows the effect of physically-motivated inner products and the correct approximation of the volume integral on the state vector norm.
The density-based state vector norms presented in figure \ref{fig:state_norms} (a) for the unweighted and weighted state vary $\approx 1.5\%$ and $\approx 0.7\%$, respectively.
The reduction due to weighting demonstrates its importance in reducing mesh dependence.
The state vector norm time sequences based on the two- and three-dimensional velocity in figures \ref{fig:state_norms} (b) and \ref{fig:state_norms} (c) are almost identical and yield normalized kinetic energy fluctuations of less than $\approx 2.5\%$ and $\approx 1.7\%$ for the unweighted and weighted versions, respectively.
The differences between the two- and three-dimensional state vector norms are minimal because the spanwise velocity is small compared to the other two components and thus contributes little to the flow's kinetic energy.
Finally, we present the normalized state vector norms based on the local speed of sound in combination with the velocity vector in figure \ref{fig:state_norms} (d).
The latter definition corresponds to the state vector in equation \eqref{eq:final_state}.
This physically-motivated state vector fluctuates less than $\approx 0.2\%$ and $\approx 0.1\%$ for the unweighted and weighted versions, respectively.
For all state vectors, weighting  significantly reduces temporal fluctuations in the state's norm. 
For UDMD, the physically-motivated state vector according to \eqref{eq:final_state} with its conserved norm is clearly the most suitable one.

\begin{figure}[htbp]
    \centering
    \includegraphics[width=0.8\textwidth]{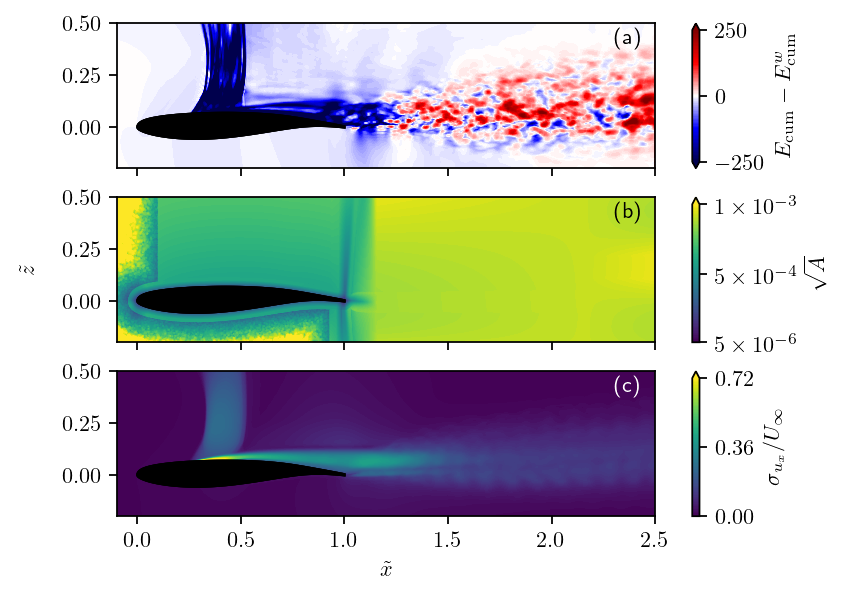}
    \caption{Mesh dependence analysis showing (a) the difference in the cumulative error \eqref{eq:err_cum} without and with weighting, (b) the square-root of the face area $\sqrt{A}$, and (c) the temporal standard deviation $\sigma_{u_x}$ normalized by the freestream velocity $U_\infty$. The state vector is based on $u_x$ and $u_z$, but only the $u_x$ component is shown.}
    \label{fig:mesh_bias}
\end{figure}

As mentioned above, the weighting of the state vector is also important to mitigate a potential influence of the computational mesh on the DMD analysis. To visualize this mesh bias, we compute the cumulative projection error defined as:
\begin{equation}
    \label{eq:err_cum}
    E_{\mathrm{cum}} = \sum\limits_{i=1}^{N_\mathrm{snap}-1} \left|\mathbf{x}_i - \mathbf{Ax}_{i-1}\right|,
\end{equation}
where $\left|\cdot\right|$ is the element-wise absolute value of the difference.
The standard definition of the DMD operator is employed and the rank is selected according to Gavish and Donoho \cite{gavish2014} (equation \eqref{eq:optRank}).
The state vector is based on the velocity components $u_x$, and $u_z$.
The cumulative error quantifies the spatial distribution of the prediction error. 
To assess the mesh dependence on the DMD analysis we show the difference in $E_{\mathrm{cum}}$ without and with weighting, i.e. $E_{\mathrm{cum}}-E_{\mathrm{cum}}^w$ in figure \ref{fig:mesh_bias} (a). 
Blue regions indicate a lower error in the unweighted state vector prediction, hence, the difference is negative. 
In contrast, red regions indicate a lower error of the weighted state vector. 
To understand this behavior, figures \ref{fig:mesh_bias} (b) and (c) show the square root of the face area $\sqrt{A}$ (the weight) and the temporal standard deviation, respectively.
The unweighted state vector yields more accurate results in the region surrounding the airfoil, where the mesh is very fine and the flow fluctuations are high.
On the other hand, the weighted state vector yields a more accurate prediction in the wake, where the mesh is less resolved and the fluctuations are smaller.
The projection error $||\mathbf{Y}-\mathbf{AX}||_F$ is almost identical for both variants, indicating that the error is only distributed differently.
 It is clear that regions of high mesh density and high temporal fluctuations dominate the least-squares problem \eqref{eq:least_squares}. 
Thus, weighting the state vector yields a more balanced distribution of the error.

These observations were also reflected in the difference in $E_{\mathrm{cum}}$ for the TDMD and UDMD variants (not shown) and the state vectors shown in figure \ref{fig:state_norms}.
A consistent mesh bias was observed when no weighting was employed.
Note that the mesh bias in the present study is rather moderate because the mesh density in the wake is high due to the tandem configuration. As the wake resolution is reduced, the mesh dependence is expected to become more significant.

%% file: rank_sensitivity.tex
In the first of a series of tests, we assess the ability of different DMD variants to robustly identify the buffet frequency under different levels of rank truncation.
Rank truncation is important for a number of reasons. 
When using DMD as an analysis tool, a smaller rank facilitates the identification of meaningful modes because the total number of modes is reduced and because the differences in $I$ between relevant and irrelevant modes typically increases with decreasing $r$. 
Moreover, rank truncation removes potential noise from the data, making the results more robust.
Note that selective sampling of the snapshots while using the full POD basis does not avoid the problem of rank sensitivity.
In this scenario, the noise sensitivity is shifted from the rank truncation to the subjective selection of the snapshots used to form the data matrix. 
If changes in $r$ affect the DMD results while the data matrix remains the same, then adding or removing a few snapshots from the data matrix while using the full POD basis will likely have a similar effect. 
The analysis becomes more robust if the rank truncation is sufficiently strong to mitigate the influence of noise.
% Note that using the full POD basis does not avoid the problem of rank sensitivity.
% What we call rank sensitivity is actually a sensitivity to noise (small-scale turbulence), because the noise inflates the observed rank of the data matrix.
% Adding or removing a few snapshots from the data matrix while using the full POD basis shows a similar sensitivity to noise as changing the rank truncation.
% The only way to reduce this sensitivity is to use a suitable rank truncation that reduces the noise.

\begin{figure}[htbp]
    \centering
    \includegraphics[width=0.8\textwidth]{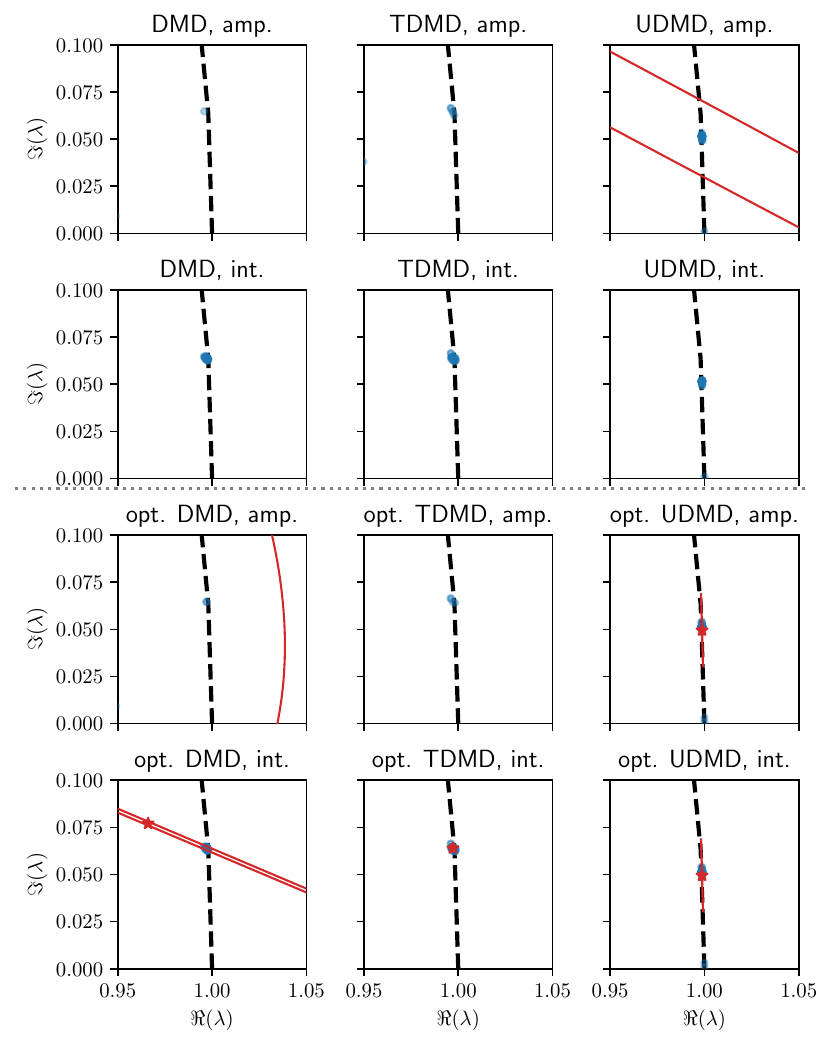}
    \caption{The blue circles mark the most dominant eigenvalues according to $I_i$ over a rank range of $5 < r < 200$. The red star indicates the mean eigenvalue over the rank range.
    The red ellipse is oriented along the two principal components of the eigenvalues, and the half-axes span twice the standard deviation along each axis.
    }
    \label{fig:eigval_rank}
\end{figure}

% The possible combinations of inner products, pre-processing, operator identification, amplitude computation, and mode selection according to figure \ref{fig:dmd_overview} are significant, even after excluding some of the more costly options.
% Therefore, we reduce the options further by assessing first the capability to identify at least the eigenvalue associated with the shock motion correctly while varying the POD basis.
Figure \ref{fig:eigval_rank} shows close-ups of the unit circle for twelve DMD variants.
Each subfigure presents the dominant eigenvalues for a rank variation spanning $5<r<200$. 
The red star indicates the mean eigenvalue over the tested rank range,
whereas the red ellipse quantifies the extent of the scatter.
The ellipse is oriented along the two principal components of the eigenvalues, with the half-axes spanning twice the standard deviation along each axis.
Subfigures with a non-visible ellipse indicate a scatter beyond the shown range.

For each value of $r$, the state vector is based on the $u_x$ and $u_z$ components of the velocity vector, and the dominant eigenvalue is identified according to two mode selection criteria.
Hence, the spectral contribution $I_i$ of the eigenvalue $i$ is either the largest magnitude of the amplitude $b_i$ (amp.) or of the integral criterion (int.).
We reasonably assume that the most dominant eigenvalue with a positive imaginary part is that of the buffet.
We recall that $r$ determines the dimensionality of the POD basis.
As $r$ increases, smaller flow structures or potential noise are included, which, ideally, should not affect the dominant buffet eigenvalue.
Therefore, the markers in figure \ref{fig:eigval_rank} should cluster in a small region as the rank is varied.
However, as the figure shows, many variants exhibit scatter.
The scatter is reduced, i.e., the robustness increases, when the mode selection is based on the integral criterion rather than the amplitude.
A further significant robustification is achieved by using optimal mode amplitudes.
We note that the dominant mode selection based on the standard amplitude is the least robust choice. 
It identifies the buffet mode only on a few occasions.
As noted by Tu et al. \cite{tu2014}, noise can lead to modes with large amplitudes and strong decay rates that contribute very little to the flow evolution but are misleading in the interpretation of the DMD spectrum.
Probably because of this sensitivity, many authors \cite{ohmichi2018,poplingher2019,hoffmann2021,masini2020} have resorted to alternative selection criteria such as the integral criterion, sparsity promotion, greedy selection or the eigenvalue weighting of Tu et al. \cite{tu2014}.

These observations hold for all other state vectors tested in this work as well.

DMD and TDMD perform similarly, but the scatter within the dominant cluster is slightly greater for TDMD. 
In comparison, UDMD delivers perfectly consistent results, especially when combined with optimal amplitudes. 
%The same is true for the other state vectors tested. 
%Only for the surface pressure (not shown) is UDMD unable to consistently identify the buffet eigenvalue, which is not surprising given the state conservation analysis presented in section \ref{sec:dmd_operator}.

\begin{figure}[t!]
    \centering
    \includegraphics[width=0.8\textwidth]{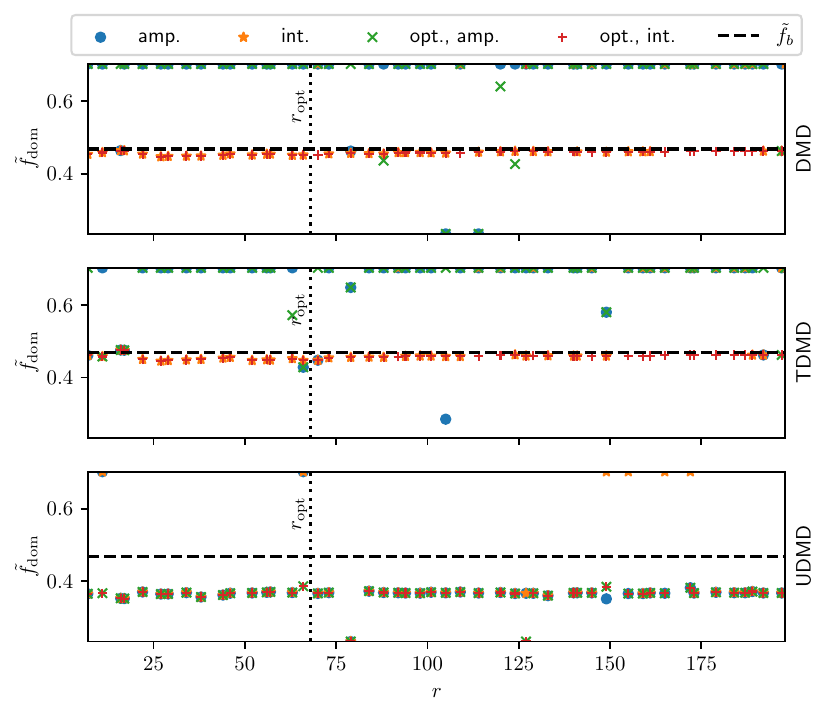}
    \caption{Dominant frequency $\tilde{f}_\mathrm{dom}$ according to $I_i$ depicted against the rank for various DMD variants. To improve visibility, markers below $0.9\tilde{f}_b$ and above $1.1\bar{f}_b$ are pinned to the plot's lower and upper boundary. $r_\mathrm{opt}$ denotes the optimal rank according to equation \eqref{eq:optRank}.}
    \label{fig:buffet_freq_rank}
\end{figure}

The above-mentioned observations are also mirrored in figure \ref{fig:buffet_freq_rank}, which presents the frequency associated with the most dominant eigenvalue according to the importance $I$ (either amplitude or integral criterion), $\tilde{f}_\mathrm{dom}$, against the rank for various DMD variants.
To improve visibility, markers with $\tilde{f}_\mathrm{dom}<0.9\tilde{f}_b$ and $\bar{f}_\mathrm{dom} > 1.19\tilde{f}_b$ are pinned to the plot's lower and upper boundaries, respectively.
The reference frequency $\tilde{f}_b$ is the normalized buffet frequency based on the autospectral density function of the lift coefficient \cite{kleinert2023}.
$r_\mathrm{opt}$ denotes the optimal rank according to equation \eqref{eq:optRank}.
%We use the buffet frequency identified from the autospectral density function of the lift coefficient as a reference \cite{kleinert2023}.
% Figure \ref{fig:eigval_rank} shows the eigenvalues' spread in the complex plane, but the link to the rank parameter is lost.
% Therefore, the identified buffet frequency $\tilde{f}_b$ is depicted against the rank in figure \ref{fig:buffet_freq_rank}.
In addition to the previous observations, figure \ref{fig:buffet_freq_rank} provides the first quantitative description of the rank sensitivity.
Both DMD and TDMD are rank-sensitive, but TDMD produces slightly more outliers and scatter around the reference frequency.
Again, the optimal mode amplitude and integral mode selection approaches exhibit the most robust and accurate results.
%Of the three operator variants, the standard DMD matches the buffet frequency best when the POD basis is sufficiently truncated, e.g., $r<400$.
For  DMD and TDMD, the optimal rank $r_\mathrm{opt}$ according to equation \eqref{eq:optRank} appears to be a good compromise between robustness and flow resolution.
UDMD is the least rank-sensitive algorithm, regardless of the mode amplitude and mode selection approach.
However, its $\tilde{f}_\mathrm{dom}$ is, on average, significantly lower than the reference $\tilde{f}_b$.
It is not entirely clear why UDMD delivers a frequency shift of the dominant mode.

Before going any further, it is important to mention another shortcoming of UDMD.
In our analysis, we observed that strange artifacts appear in the buffet mode distribution (as well as other modes) when the rank is changed.
The DMD buffet mode remains almost identical over a wide range of rank truncations.
Again, it is not entirely clear why UDMD exhibits these artifacts.
We refer the reader to appendix \ref{sec:dmd_vs_udmd} for a comparison between the DMD and UDMD mode distributions.
For these reasons, we mostly employ henceforth the standard DMD variant for operator selection.

\begin{figure}[htbp]
    \centering
    \includegraphics[width=0.8\textwidth]{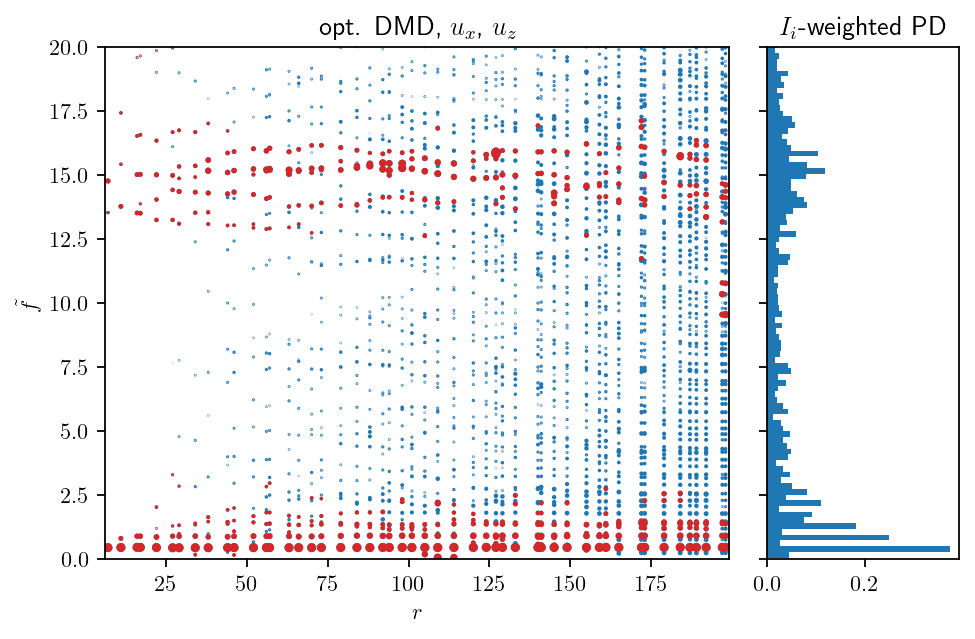}
    \caption{DMD spectra over the rank range $5<r<200$. The marker size is scaled by $I_i$ (integral criterion) and the top 10 markers are colored in red.}
    \label{fig:rank_spectrum}
\end{figure}

The rank sensitivity investigation is not complete without examining the full spectrum.
% Next, we investigate the rank sensitivity of the full spectrum. 
Figure \ref{fig:rank_spectrum} shows full DMD spectra for ranks $5 < r < 200$ using the optimal DMD variant with integral mode selection.
The size of the markers is scaled by the integral criterion,
and the ten most dominant modes are colored red.
The probability density (PD) to the right of the scatter plot visualizes the regularity of occurrence of each frequency.
Ideally, the red markers follow a straight horizontal line, which corresponds to a low rank sensitivity of the dominant modes.
As can be seen, this robust DMD variant consistently identifies not only the buffet frequency but also several of its harmonics over a wide range of ranks.
For higher frequencies, the variability increases with the rank but remains acceptable.
It is worth noting the identified important modes at $\tilde{f}\approx 15$, which we shall revisit later in the manuscript.

% The rank sensitivity is also influenced by the choice of the state vector.
% Figure \ref{fig:rank_spectrum_rho} shows DMD spectra for the density-based state vector.
% Compared to figure \ref{fig:rank_spectrum_uxya}, the rank sensitivity is much larger, especially for frequencies $\tilde{f} > 2$.
% Moreover, no relevant modes in the vortex shedding regime about $\tilde{f}\approx 10$ are detected.
% Thus, physically-motivated state vectors are better suited for the DMD analysis, even if the unitary DMD variant is not employed.

%\begin{figure}[t!]
%    \centering
%    \includegraphics[width=0.8\textwidth]{figures/f_over_rank_rho.png}
%    \caption{DMD spectra for $5<r<250$ and the density-based state vector. The marker size is scaled by $I_i$ (integral criterion) and the top 10 markers are colored in red.}
%    \label{fig:rank_spectrum_rho}
%\end{figure}

% Considering the full spectrum, both DMD and UDMD deliver very similar results in the investigated range of $r$.
% We did not consider the TDMD in this test due to the worse performance in the previous tests.

%% file: sampling_rate_sensitivity.tex
In the second set of tests we investigate the sensitivity of DMD and TDMD with optimal amplitudes to the sampling rate $\tilde{f}_s$.
The rank is chosen automatically following Gavish and Donoho \cite{gavish2014}.
The state vector is still based on the weighted $u_x$ and $u_z$ velocity components.
We gradually decrease the sampling rate from the maximum value $\tilde{f}_s/\tilde{f}_b =200$, halving the number of snapshots at each step.

\begin{figure}[t!]
    \centering
    \includegraphics[width=0.8\textwidth]{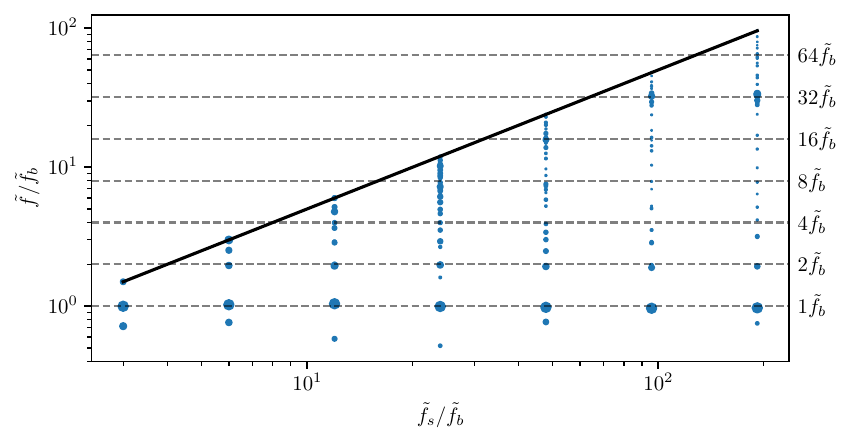}
    \caption{Spectra computed with the opt. DMD for a range of sampling rates. The solid line demarcates the Nyquist frequency. The marker size is scaled by the integral criterion.}
    \label{fig:topk_freq_sampling}
\end{figure}

\begin{figure}[t!]
    \centering
    \includegraphics[width=0.8\textwidth]{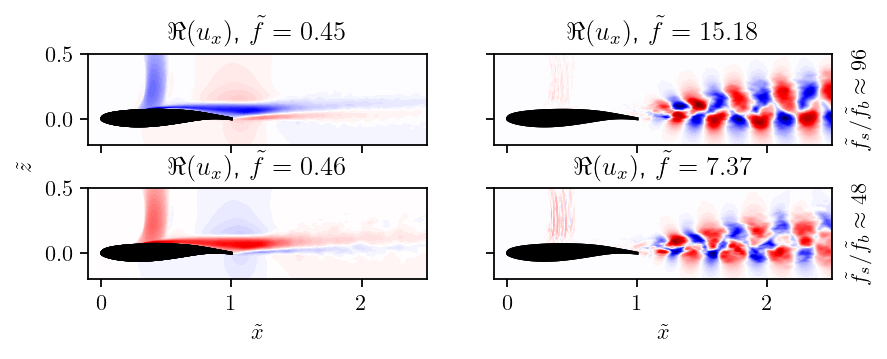}
    \caption{Dependency of two most dominant DMD mode distributions and associated frequencies on the sampling frequency; mode selection based on the integral criterion.}
    \label{fig:fs_mode_dep}
\end{figure}

Figure \ref{fig:topk_freq_sampling} presents the opt. DMD spectra for a range of sampling rates.
The results for the opt. TDMD are nearly identical, which is why we do not include them here.
The marker size is scaled by the integral criterion, and the solid line demarcates the Nyquist frequency.
Examining the markers near the horizontal line at $\tilde{f}/\tilde{f}_b = 1$ clearly shows that the buffet frequency is always reliably identified, even if only three snapshots per buffet cycle are used.
However, for sampling rates below $\tilde{f}_s/\tilde{f}_b \approx 100$, the upper end of the spectrum gets clipped.
The same threshold value was also reported by Poplingher et al. \cite{poplingher2019} in their DMD analysis.
This clipping is associated with aliasing, where high-frequency modes associated with wake phenomena, are now aliased to lower frequencies.
This effect is observed by the frequency clustering around $f_\mathrm{Nyquist}$ and by the high-frequency modes now appearing at lower frequencies, as illustrated in figure \ref{fig:fs_mode_dep}.
The high-frequency mode identified at $\tilde{f}=15.18$ with a sampling rate of $\tilde{f}_s/\tilde{f}_b =96$ erroneously appears at about half its frequency ($\tilde{f}=7.37$) with a sampling rate of $\tilde{f}_s/\tilde{f}_b=48$.
Other similar mode-folding at different frequencies (not shown) is also observed. 
The buffet mode with its relatively low frequency remains unaffected.

We note the difficulty of avoiding such aliasing in experimental settings.
Considering the values listed in table \ref{tab:data_sources} and the state-of-the-art field measurement techniques, the currently achievable sampling rates are barely sufficient to capture the shedding frequency.

%% file: slice_analysis.tex
Based on the test results, we employ the following DMD settings in this final application:
\begin{itemize}
    \item \textbf{State vector:} the state vector in the first part of the analysis is based on $u_x$ and $u_z$ and is weighted by the cell volume/area. To evaluate the propagation speed of the pressure close to the airfoil's surface, we employ a state based on the pressure $p$ weighted by the cell volume. Every second snapshot is used, giving a total of 200 snapshots. The corresponding sampling frequency allows detecting frequencies up to $\tilde{f}_{\mathrm{max}}\approx 22$.
    \item \textbf{Operator:} the standard operator definition \eqref{eq:A_def} is used without unitary constraint. 
    \item \textbf{Rank: } the size of the SVD basis is chosen according to equation \eqref{eq:optRank}.
    \item \textbf{Mode amplitudes:} the optimization problem \eqref{eq:opt_dmd} is solved to obtain the \textit{optimal} mode amplitudes.
    \item \textbf{Mode selection:} the importance of the modes is evaluated according to the integral criterion \eqref{eq:int_criterion}.
\end{itemize}
Similarly successful implementations on surface and volumetric data of a NACA 0012 airfoil using the same settings are provided in the supplementary repository.

Figure \ref{fig:slice_dmd_freq} presents the resulting DMD spectrum that exhibits a clear peak at $\tilde{f}_b\approx 0.46$, which is in agreement with the frequency extracted from the lift coefficient and the values reported by Jacquin et al. \cite{jacquin2009} and Szubert et al. \cite{szubert2015}. The second largest peak occurs at $\tilde{f}\approx 15.2$, which is also in excellent agreement with the literature \cite{kleinert2023,szubert2015}. No dominant modes were detected beyond the vortex shedding frequency. Besides the buffet frequency, two of its harmonics also appear in the spectrum, which is aligned with Poplingher et al. \cite{poplingher2019}.

% As can be seen in the animations in the supplementary material, the vortex shedding is not regular but rather starts slowly, reaches a maximum shedding frequency, and then ceases within a buffet cycle.
% Consequently, there is not a unique vortex shedding frequency.

\begin{figure}[t!]
    \centering
    \includegraphics[width=0.8\textwidth]{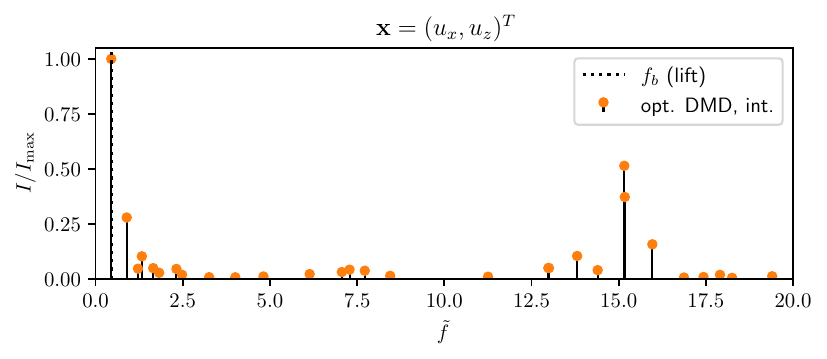}
    \caption{DMD spectrum. The dashed line denotes the buffet frequency $\tilde{f}_b\approx 0.46$ based on the lift coefficient.}
    \label{fig:slice_dmd_freq}
\end{figure}

\begin{figure}[t!]
    \centering
    \includegraphics[width=0.8\textwidth]{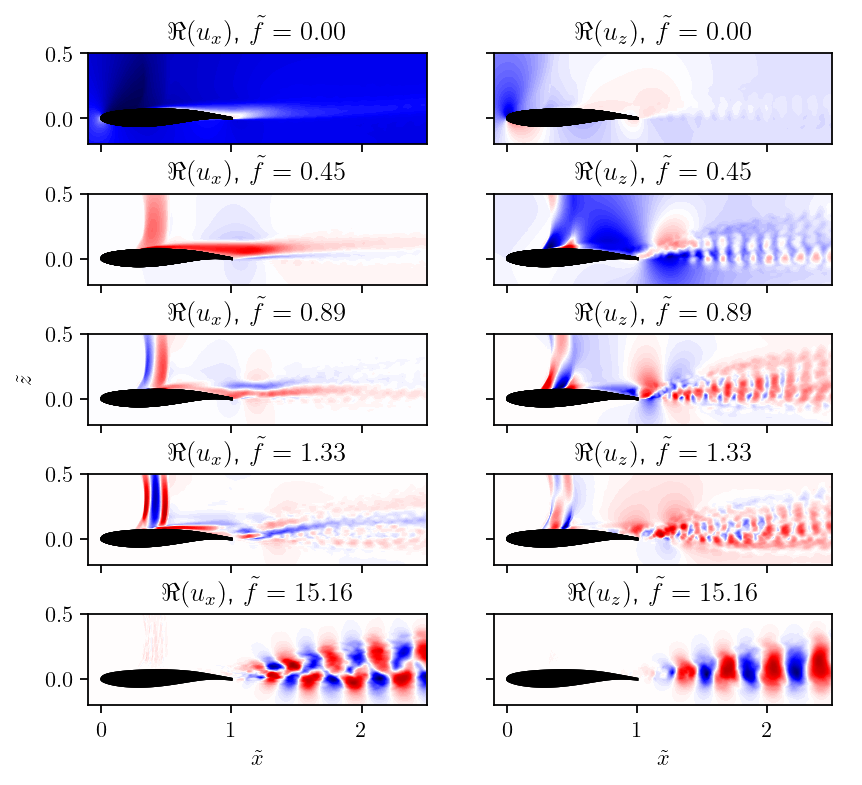}
    \caption{Real part of the top five DMD modes. The figures only show the real part.}
    \label{fig:slice_dmd_modes}
\end{figure}

Figure \ref{fig:slice_dmd_modes} presents the real part of the five most dormant modes sorted by frequency: the steady background mode, the buffet mode and two of its harmonics, and the vortex shedding mode.
The $u_x$-component of the buffet mode shows activity about the shock, within the separated boundary layer, and in the rear part above the boundary layer, which is in agreement with Das et al. \cite{das2022}. All modes are qualitatively similar to the POD modes reported by Kleinert et al. \cite{kleinert2023} and Szubert et al. \cite{szubert2015}. To quantify how much of the flow dynamics these five modes encode, we define the reconstruction error as:
\begin{equation}
    \label{eq:rec_err}
    E = ||\mathbf{M}-\mathbf{\Phi D}_\mathbf{b} \mathbf{V}_\mathbf{\omega}||_F,
\end{equation}
where $\mathbf{M}$ is the full snapshot matrix, $\mathbf{D}_\mathbf{b}$ is a diagonal matrix based on the amplitudes, and $\mathbf{V}_\mathbf{\omega}$ is the Vandermode matrix. To have a more expressive version of the error, we normalize it as:
\begin{equation}
    \label{eq:rec_err_norm}
     \tilde{E} = \frac{E - E_\mathrm{min}}{E_\mathrm{max}-E_\mathrm{min}},
\end{equation}
where $E_\mathrm{min}$ is the reconstruction error using all modes and $E_\mathrm{max}$ is the reconstruction error using only the background mode, i.e. all elements of $\mathbf{D}_\mathbf{b}$ are set to zero except for the one corresponding to the background mode.
Considering the normalized error \eqref{eq:rec_err_norm}, the first five modes and their conjugate complex counterparts according to the greedy selection algorithm of Ohmichi et al. \cite{ohmichi2018} reduce the prediction error by $\approx 72\%$.
Of this reduction, the buffet mode contributes $\approx 62\%$ and the shedding mode an additional $\approx 5\%$ (considering the drop in $\tilde{E}$ when these modes are used in isolation).

Compared to the greedy algorithm, the integral criterion yields very similar top five modes in a slightly different order. 
The greedy algorithm provides a smoother decay of $\tilde{E}$. 
However, considering the first five modes in combination, the integral criterion reduces $\tilde{E}$ by $\approx 81\%$, whereas the greedy selection yields a reduction of only $72\%$. 
Note that these observations are most likely case-dependent and do not generalize. 
Since the integral criterion is significantly cheaper to compute, we prefer it over the greedy selection, at least when the focus of the DMD is on flow analysis.

%Downstream of the shock, the $u_y$-component exhibits a region of high values where the $u_x$-component has a minimum, indicating a large periodic displacement of fluid normal to the airfoil's surface.
%The $a$-component of the buffet mode appears mostly like a mixture of the other two components.

% The second mode at $\tilde{f}=12.58$ captures flow separation in the boundary layer and the wake.
% The structures with alternating intensity correspond to the expected vortex shedding.
% Moreover, vertical lamellar structures below and above the airfoil can be observed, particularly in the $u_x$- and $a$-components.
The presented modes in figure \ref{fig:slice_dmd_modes} exhibit the typical distributions associated with the buffet (and its harmonics) and a shedding mode. 
Unlike the NACA 0012 analysis in the online repository, the vortex shedding mode at $\tilde{f}= 15.16$ does not clearly show the expected acoustic waves.
These are better visualized by the phase angle distribution, defined as:
\begin{equation}
    \phi=\tan^{-1}\left(\frac{\Im(\varphi}){\Re(\varphi)}\right)\,
\end{equation}
where $\Im(\varphi)$ and $\Re(\varphi)$ denote the imaginary and real components of the mode $\varphi$.

%The nature of these waves is elucidated through the examination of the phase angle and the wave propagation speed.
%Figure \ref{fig:slice_dmd_phase} presents the phase angle distribution,
%which is computed as
%The white contour lines demark the phase at $\phi = 0$.
% seen before and also small-scale disturbances in the shock region.
The direction of wave propagation can be deduced from the phase angle gradient \cite{poplingher2019}.
Outside the wake and above and below the airfoil, the $\tilde{x}$ slope is positive, indicating that the waves are traveling upstream.
On the other hand, the phase angle slope in $\tilde{x}$ inside the separated wake is negative, indicating downstream wave propagation.

\begin{figure}[t!]
    \centering
    \includegraphics[width=0.8\textwidth]{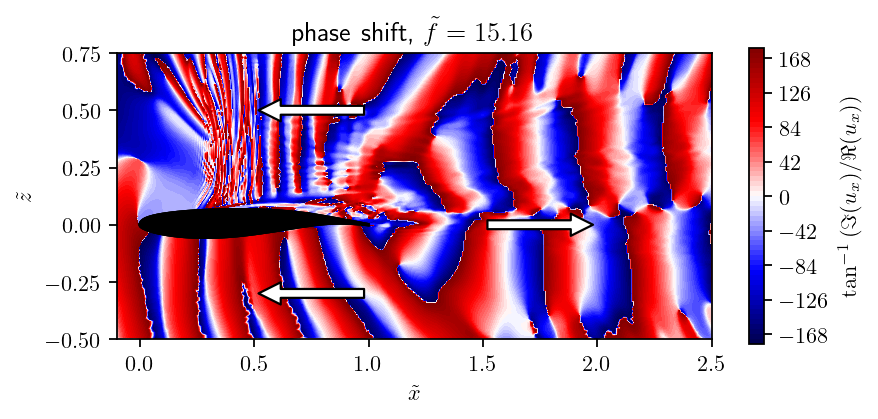}
    \caption{Phase angle distribution of a shedding mode.}
    \label{fig:slice_dmd_phase}
\end{figure}

Besides the propagation direction, the phase angle gradient can also be used to compute the wave propagation speed in some direction $s$ \cite{poplingher2019}:
\begin{equation}
    U_p = \frac{2\pi f}{\Delta\phi / \Delta s}.
\end{equation}
%However, in this study, we opt for using the spatio-temporal correlation coefficient  \cite{hoffmann2021} to infer the wave speed along a line.
%Figure \ref{fig:mode_correlation} presents the spatio-temporal correlation coefficient of the velocity magnitude from DMD-reconstructed snapshots using a single mode at $\tilde{f}=12.58$.
%The correlation is evaluated along $0.7 \le \tilde{y} \le 0.8$ at $\tilde{x}=0.5$.

\begin{figure}[t!]
    \centering
    \includegraphics[width=0.8\textwidth]{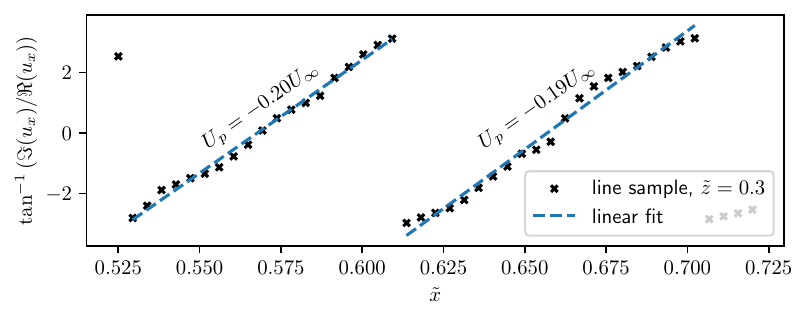}
    \caption{Phase angle variation along the line $0.52 \leq \tilde{x} \leq 0.72$ and $\tilde{z}=0.3$ of the shedding mode.}
    \label{fig:phase_speed}
\end{figure}

%The propagation velocity can be elucidated from the slope of the ridge lines \cite{hoffmann2021}.
Figure \ref{fig:phase_speed} shows the propagation speed of the shedding mode evaluated along the line $0.52 \leq \tilde{x} \leq 0.72$ and $\tilde{z}=0.3$.
The waves propagate upstream at a speed of $U_p=0.2U_\infty$.
This value is lower than the $0.36U_\infty$ reported by Hoffmann et al. \cite{hoffmann2021} based on their PIV measurements of the same airfoil at $Re=1.9\times 10^6$, $Ma_\infty = 0.72$, and $\alpha=3.5^\circ$.
Reasons for the observed difference may be the higher angle of attack in the present study and the presence of numerical dissipation, which dampens the acoustic waves.

%As expected, the velocity is negative (i.e., pointing upstream) and very close to the speed of sound in the absolute frame of reference (the free stream velocity is set such that $M_\infty = 0.75$).
%Hence, these waves at $\tilde{f}=12.58$ are the well-known acoustic waves that originate in the shear layer and the trailing edge and propagate upstream \cite{lee1990,giannelis2017}.
%They are clearly \textcolor{blue}{locked} to the shear layer dynamics.

%\begin{figure}[t!]
%    \centering
%    \includegraphics[width=0.8\textwidth]{figures/dmd_mode_line_correlation_mode_256.png}
%    \caption{Spatio-temporal correlation coefficient of the velocity magnitude from DMD reconstructed snapshots using a single mode at $\tilde{f}=12.58$. The correlation is evaluated along $0.7 \le \tilde{y} \le 0.8$ at $\tilde{x}=0.5$. The dashed line demarks the wave propagation speed.}
%    \label{fig:mode_correlation}
%\end{figure}

\begin{figure}[t!]
    \centering
    \includegraphics[width=0.8\textwidth]{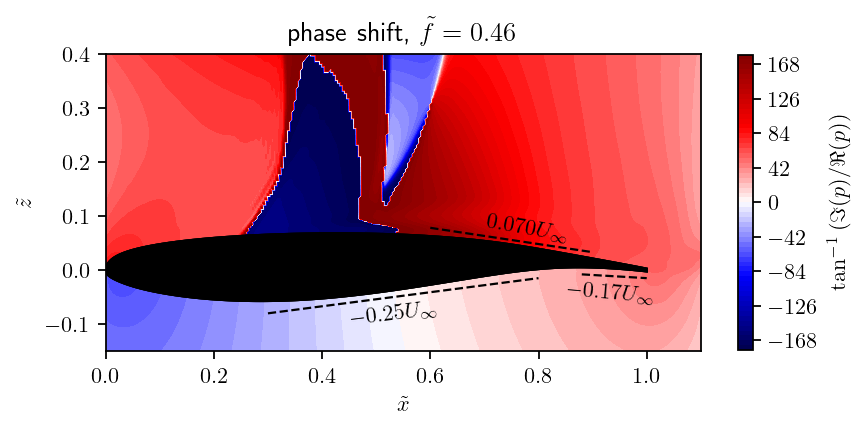}
    \caption{Phase angle distribution of buffet mode; the state vector is based on the pressure field $p$; the propagation speed of pressure waves is evaluated along the dashed lines; a negative sign indicates an upstream propagation.}
    \label{fig:slice_dmd_phase_buffet}
\end{figure}

Besides the propagation speed of acoustic waves above the airfoil, it is also beneficial to compute the propagation speed of surface pressure waves.
For this last part of the analysis, we switch from the velocity-based to the more sensible pressure state vector, because of the large velocity gradient near the surface.
The remaining DMD settings are exactly the same as before (refer to the online repository).
Qualitatively, the resulting DMD spectrum is very close to the one shown in figure \ref{fig:slice_dmd_freq}. 
Only the relative importance of the vortex shedding mode compared to the buffet mode decreases slightly when the pressure is analyzed.
As can be inferred from figure \ref{fig:slice_dmd_modes}, the buffet at $\tilde{f}=0.46$ dominates the flow evolution on the airfoil's suction side, which is also true for the pressure-based results.
Therefore, we extract the propagation speed from the buffet mode as performed by Poplingher et al. \cite{poplingher2019}.
Figure \ref{fig:slice_dmd_phase_buffet} shows an enlarged view of the phase angle distribution around the airfoil.
The figure also shows propagation speeds along several sample lines near the surface.
On the suction side in the range $0.6 \le \tilde{x} \le 0.9$, the pressure propagates downstream at a nearly constant speed of $0.070 U_\infty$. This value is in close agreement with that reported by Jacquin et al. \cite{jacquin2009} ($0.0718U_\infty$) and Poplingher et al. \cite{poplingher2019} ($0.0826U_\infty$).
On the pressure side, the waves propagate in upstream direction. In the range $0.3\le \tilde{x} \le 0.8$, the speed of $-0.25U_\infty$ is again very close to the $-0.271U_\infty$ and $0.2837U_\infty$ reported in \cite{jacquin2009} and \cite{poplingher2019}, respectively.
Finally, also the propagation speed on the pressure side near the trailing edge agrees with the value reported by Poplingher et al. in the same region ($0.1651U_\infty$).

Before concluding this section, we would like to point out the simplicity of computing the wave propagation speed.
While Feldhusen-Hoffmann et al. \cite{hoffmann2021} applied a bandpass filter to the PIV data before computing the spatio-temporal correlation, such filtering is not necessary for a DMD mode since it is associated with a single frequency.
Omitting the filtering step is a clear advantage since no prior knowledge of the wave frequency range is required.

% Next, we want to emphasize the importance of DMD as the modal decomposition technique for investigating shock buffets.
% Figure \ref{fig:slice_svd_modes} shows the leading POD modes related to shock buffet and vortex shedding.
% While both POD modes look very similar to the corresponding DMD modes displayed in figure \ref{fig:slice_dmd_modes}, the POD vortex shedding mode misses the sound waves due to their low energy content.
% Instead, the DMD modes reliably establish a connection between vortex shedding and the upstream propagating waves.

% \begin{figure}[htbp]
%     \centering
%     \includegraphics[width=0.8\textwidth]{figures/svd_modes_uxya_article.png}
%     \caption{Selected POD modes associated with shock buffet and vortex shedding}
%     \label{fig:slice_svd_modes}
% \end{figure}

%% file: summary.tex
In this study, we benchmark various DMD variants on simulation data of transonic shock buffet flow with the aim of identifying a robust and accurate methodology.
The analysis is enabled by the introduction of a simple modular framework of the DMD algorithm. 
The framework consists of five distinct steps,
where each step involves various choices leading to different DMD variants.

%The analysis is conducted on a scale-resolving IDDES simulation of a NACA-0012 under transonic buffet conditions.
%The IDDES simulation with the Spalart-Almaras turbulence model is performed using \textit{rhoCentralFoam} solver from OpenFOAM-v2012.
%The simulation is validated against the experimental results of McDevitt and Okuna \cite{mcdevitt1985}.

The conducted tests examine the sensitivity of different DMD variants to rank truncation and sampling rate. 
The tests yield a DMD variant with corresponding settings that provides robust and accurate identification of the relevant flow dynamics.
The suggested workflow can reliably detect the shock-boundary-layer interaction and also identifies upstream-propagating acoustic waves originating in the shear layer and at the trailing edge.
The waves' frequency is directly linked to the vortex shedding at the trailing edge.
Similar results in terms of the DMD spectra and flow phenomena encoded in the modes could be obtained based on various state vectors, which is in agreement with the observation by Ohmichi et al. \cite{ohmichi2018} and supports the possibility of comparing numerical and experimental data from different sources.

Another objective of this study is to provide recommendations on how best to perform dynamic mode decomposition on a transonic buffet flow. 
Based on the test results, the following general recommendations are derived:
\begin{enumerate}
    \item An extended, physically-motivated state vector is the safer and often better choice, especially when the projection error $||\mathbf{Y}-\mathbf{AX}||_F$ is large.
    The extended state vector according to equation \eqref{eq:final_state} can improve the fulfilment of the assumed linearity of the dynamical system without introducing a subjective re-scaling of the flow fields, e.g., by using min-max-scaling.
    Furthermore, physically-motivated state vectors allow the use of constrained linear operators.
    An example we tested in this work is the so-called unitary DMD, which preserves the norm/energy of the state.
    The extended state can also be useful in the construction of reduced-order models, such as DMD-based Galerkin models.
    For the purpose of flow analysis of the OAT15A data investigated here, the extended state vector did not provide any advantage over state vectors based on the velocity components.
    However, more complex flow dynamics or noisier snapshots, e.g., less regular buffet cycles or experimental data, may yield a different picture.
    Clearly, physically-motivated states are more challenging to design for experimental data, where only subsets of velocity or pressure fields are acquired.
    
    \item Weigh the cell-centered values by the square root of the cell volume (or surface area in the case of two dimensions). 
    Weighing improves the calculation of the inner product operator and reduces mesh dependence.
    \item Special care should be taken when truncating the SVD,
    which has shown the greatest sensitivity to the DMD results. 
    An automatic rank truncation according to equation \eqref{eq:optRank} was deemed appropriate for the current flow. 
    However, this criterion may not be ideal for other flows or data types. 
    It is therefore recommended that the rank sensitivity be re-examined for each new problem.
    The rank should be chosen based on an iterative process that produces a robust and accurate outcome.
    We caution against using a strong truncation that filters out the low energy-containing acoustic waves.
    \item Sample your results at a sufficiently high rate to avoid aliasing. 
    Following the Nyquist criteria, the sampling rate should be two to three times the maximum frequency of interest. 
    %\item While for a robust DMD variant and settings the time window has little influence, a reasonable time sequence should be nonetheless employed.
    %A minimum of two buffet cycles should be used for the analysis, where the time window spans an integer number of cycles.
    %For a short sequence, time windows that span a non-integer number of cycles risk skewing the results.
    \item We recommend using the standard operator definition according to equation \eqref{eq:A_def}. 
    This definition is computationally tractable and gave consistent results in our study.
    % The UDMD accurately identifies the dominant frequency and the corresponding mode occasionally exhibits strange artifacts.
    % We posit UDMD as a very promising variant that requires further investigations to uncover its full potential.
    \item The optimal mode amplitude (equation \eqref{eq:opt_dmd}) should be employed.
    It is more accurate than the standard definition and less expensive than the sparsity-promoting approach.
    \item Use the integral criterion according to equation \eqref{eq:int_criterion} to select the modes.
    This approach is best suited to handle quickly-vanishing modes that contribute very little to the full reconstruction, and is computationally more tractable than greedy mode selection, while providing comparable results.
\end{enumerate}
The provided workflow, including all routines for data processing, modal analysis, and visualization, is fully available in a supplementary repository.
%All DMD analyses are conducted with the python library flowTorch \cite{weiner2021}.

We hypothesize that this recipe is also applicable to other fluid flow problems.
The authors are currently investigating this postulate.
Other possible future investigations include the exploration of other DMD variants that `wrap' around the five-step framework we have introduced. 
Examples include bagging DMD \cite{sashidhar2022}, multiscale DMD \cite{dylewsky2019}, or higher-order DMD \cite{clainche2017}. 
Higher-order DMD could better capture the nonlinearities in the data while keeping the memory requirements tractable.
However, these `wrapper' algorithms are also more computationally demanding and deliver results that are more complex to interpret.

%% file: main.bbl
\begin{thebibliography}{42}
\newcommand{\enquote}[1]{``#1''}
\providecommand{\natexlab}[1]{#1}
\providecommand{\url}[1]{\texttt{#1}}
\providecommand{\urlprefix}{URL }
\expandafter\ifx\csname urlstyle\endcsname\relax
  \providecommand{\doi}[1]{\discretionary{}{}{}https://doi.org/#1}\else
  \providecommand{\doi}[1]{\discretionary{}{}{}\urlstyle{rm}\url{https://doi.org/#1}}\fi

\bibitem[{Giannelis et~al.(2017)Giannelis, Vio, and Levinski}]{giannelis2017}
Giannelis, N.~F., Vio, G.~A., and Levinski, O., \enquote{A review of recent
  developments in the understanding of transonic shock buffet,} \emph{Progress
  in Aerospace Sciences}, Vol.~92, 2017, pp. 39--84.
\newblock \doi{10.1016/j.paerosci.2017.05.004}.

\bibitem[{Feldhusen-Hoffmann et~al.(2021)Feldhusen-Hoffmann, Lagemann, Loosen,
  Meysonnat, Klaas, and Schr{\"o}der}]{hoffmann2021}
Feldhusen-Hoffmann, A., Lagemann, C., Loosen, S., Meysonnat, P., Klaas, M., and
  Schr{\"o}der, W., \enquote{Analysis of transonic buffet using dynamic mode
  decomposition,} \emph{Experiments in Fluids}, Vol.~62, No.~4, 2021, p.~66.
\newblock \doi{10.1007/s00348-020-03111-5}.

\bibitem[{Jacquin et~al.(2009)Jacquin, Molton, Deck, Maury, and
  Soulevant}]{jacquin2009}
Jacquin, L., Molton, P., Deck, S., Maury, B., and Soulevant, D.,
  \enquote{Experimental Study of Shock Oscillation over a Transonic
  Supercritical Profile,} \emph{AIAA Journal}, Vol.~47, No.~9, 2009, pp.
  1985--1994.
\newblock \doi{10.2514/1.30190}.

\bibitem[{Szubert et~al.(2015)Szubert, Grossi, {Jimenez Garcia}, Hoarau, Hunt,
  and Braza}]{szubert2015}
Szubert, D., Grossi, F., {Jimenez Garcia}, A., Hoarau, Y., Hunt, J.~C., and
  Braza, M., \enquote{Shock-vortex shear-layer interaction in the transonic
  flow around a supercritical airfoil at high Reynolds number in buffet
  conditions,} \emph{Journal of Fluids and Structures}, Vol.~55, 2015, pp.
  276--302.
\newblock \doi{10.1016/j.jfluidstructs.2015.03.005}.

\bibitem[{CROUCH et~al.(2009)CROUCH, GARBARUK, MAGIDOV, and
  TRAVIN}]{crouch2009}
CROUCH, J.~D., GARBARUK, A., MAGIDOV, D., and TRAVIN, A., \enquote{Origin of
  transonic buffet on aerofoils,} \emph{Journal of Fluid Mechanics}, Vol. 628,
  2009, p. 357–369.
\newblock \doi{10.1017/S0022112009006673}.

\bibitem[{Poplingher and Raveh(2023)}]{poplingher2023}
Poplingher, L., and Raveh, D.~E., \enquote{Comparative Modal Study of the
  Two-Dimensional and Three-Dimensional Transonic Shock Buffet,} \emph{AIAA
  Journal}, Vol.~61, No.~1, 2023, pp. 125--144.

\bibitem[{Ohmichi et~al.(2018)Ohmichi, Ishida, and Hashimoto}]{ohmichi2018}
Ohmichi, Y., Ishida, T., and Hashimoto, A., \enquote{Modal Decomposition
  Analysis of Three-Dimensional Transonic Buffet Phenomenon on a Swept Wing,}
  \emph{AIAA Journal}, Vol.~56, No.~10, 2018, pp. 3938--3950.
\newblock \doi{10.2514/1.J056855}.

\bibitem[{Poplingher et~al.(2019)Poplingher, Raveh, and
  Dowell}]{poplingher2019}
Poplingher, L., Raveh, D.~E., and Dowell, E.~H., \enquote{Modal Analysis of
  Transonic Shock Buffet on 2D Airfoil,} \emph{AIAA Journal}, Vol.~57, No.~7,
  2019, pp. 2851--2866.
\newblock \doi{10.2514/1.J057893}.

\bibitem[{Das et~al.(2020)Das, Carrese, Marzocca, and Levinski}]{das2020}
Das, A., Carrese, R., Marzocca, P., and Levinski, O., \emph{Flow Diagnostics of
  Transonic Shock Buffet Phenomenon of a Supercritical Airfoil using Dynamic
  Mode Decomposition}, 2020.
\newblock \doi{10.2514/6.2020-1988}.

\bibitem[{Das et~al.(2022)Das, Marzocca, Levinski, and Das}]{das2022}
Das, A., Marzocca, P., Levinski, O., and Das, R., \emph{Investigation onto Deep
  Transonic Buffet Condition of a Supercritical Airfoil using Multiresolution
  Dynamic Mode Decomposition}, 2022.
\newblock \doi{10.2514/6.2022-1956}.

\bibitem[{Masini et~al.(2020)Masini, Timme, and Peace}]{masini2020}
Masini, L., Timme, S., and Peace, A.~J., \enquote{Analysis of a civil aircraft
  wing transonic shock buffet experiment,} \emph{Journal of Fluid Mechanics},
  Vol. 884, 2020, p.~A1.
\newblock \doi{10.1017/jfm.2019.906}.

\bibitem[{Zhao et~al.(2020)Zhao, Dai, Tian, and Xiong}]{zhao2020}
Zhao, Y., Dai, Z., Tian, Y., and Xiong, Y., \enquote{Flow characteristics
  around airfoils near transonic buffet onset conditions,} \emph{Chinese
  Journal of Aeronautics}, Vol.~33, No.~5, 2020, pp. 1405--1420.
\newblock \doi{10.1016/j.cja.2019.12.022}.

\bibitem[{Schmid(2010)}]{schmid2010}
Schmid, P.~J., \enquote{Dynamic mode decomposition of numerical and
  experimental data,} \emph{Journal of Fluid Mechanics}, Vol. 656, 2010, p.
  5–28.
\newblock \doi{10.1017/S0022112010001217}.

\bibitem[{Kutz et~al.(2016{\natexlab{a}})Kutz, Brunton, Brunton, and
  Proctor}]{kutz2014}
Kutz, J.~N., Brunton, S.~L., Brunton, B.~W., and Proctor, J.~L., \emph{Dynamic
  Mode Decomposition}, Society for Industrial and Applied Mathematics,
  Philadelphia, PA, 2016{\natexlab{a}}.
\newblock \doi{10.1137/1.9781611974508}.

\bibitem[{Schmid et~al.(2011)Schmid, Li, Juniper, and Pust}]{schmid2011}
Schmid, P.~J., Li, L., Juniper, M.~P., and Pust, O., \enquote{Applications of
  the dynamic mode decomposition,} \emph{Theoretical and Computational Fluid
  Dynamics}, Vol.~25, No.~1, 2011, pp. 249--259.
\newblock \doi{10.1007/s00162-010-0203-9}.

\bibitem[{Wu et~al.(2021)Wu, Brunton, and Revzen}]{wu2021}
Wu, Z., Brunton, S.~L., and Revzen, S., \enquote{Challenges in dynamic mode
  decomposition,} \emph{Journal of The Royal Society Interface}, Vol.~18, No.
  185, 2021, p. 20210686.
\newblock \doi{10.1098/rsif.2021.0686}.

\bibitem[{Schmid(2022)}]{schmid2022}
Schmid, P.~J., \enquote{Dynamic Mode Decomposition and Its Variants,}
  \emph{Annual Review of Fluid Mechanics}, Vol.~54, No.~1, 2022, pp. 225--254.
\newblock \doi{10.1146/annurev-fluid-030121-015835}.

\bibitem[{Jovanović et~al.(2014)Jovanović, Schmid, and
  Nichols}]{jovanovic2014}
Jovanović, M.~R., Schmid, P.~J., and Nichols, J.~W.,
  \enquote{Sparsity-promoting dynamic mode decomposition,} \emph{Physics of
  Fluids}, Vol.~26, No.~2, 2014, p. 024103.
\newblock \doi{10.1063/1.4863670}.

\bibitem[{Ohmichi(2017)}]{ohmichi2017}
Ohmichi, Y., \enquote{Preconditioned dynamic mode decomposition and mode
  selection algorithms for large datasets using incremental proper orthogonal
  decomposition,} \emph{AIP Advances}, Vol.~7, No.~7, 2017, p. 075318.
\newblock \doi{10.1063/1.4996024}.

\bibitem[{Kou and Zhang(2017)}]{kou2017}
Kou, J., and Zhang, W., \enquote{An improved criterion to select dominant modes
  from dynamic mode decomposition,} \emph{European Journal of Mechanics -
  B/Fluids}, Vol.~62, 2017, pp. 109--129.
\newblock \doi{10.1016/j.euromechflu.2016.11.015}.

\bibitem[{Tu et~al.(2014)Tu, Rowley, Luchtenburg, Brunton, and Kutz}]{tu2014}
Tu, J.~H., Rowley, C.~W., Luchtenburg, D.~M., Brunton, S.~L., and Kutz, J.~N.,
  \enquote{On dynamic mode decomposition: Theory and applications,}
  \emph{Journal of Computational Dynamics}, Vol.~1, No.~2, 2014, pp. 391--421.
\newblock \doi{10.3934/jcd.2014.1.391}.

\bibitem[{Kutz et~al.(2016{\natexlab{b}})Kutz, Fu, and Brunton}]{kutz2016}
Kutz, J.~N., Fu, X., and Brunton, S.~L., \enquote{Multiresolution Dynamic Mode
  Decomposition,} \emph{SIAM Journal on Applied Dynamical Systems}, Vol.~15,
  No.~2, 2016{\natexlab{b}}, pp. 713--735.
\newblock \doi{10.1137/15M1023543}.

\bibitem[{Weiner and Semaan(2022)}]{weiner2022}
Weiner, A., and Semaan, R., \emph{Simulation and modal analysis of transonic
  shock buffets on a NACA-0012 airfoil}, 2022.
\newblock \doi{10.2514/6.2022-2591}.

\bibitem[{Brunton and Kutz(2019)}]{brunton2019}
Brunton, S.~L., and Kutz, J.~N., \emph{Data-Driven Science and Engineering:
  Machine Learning, Dynamical Systems, and Control}, Cambridge University
  Press, 2019.
\newblock \doi{10.1017/9781108380690}.

\bibitem[{Gavish and Donoho(2014)}]{gavish2014}
Gavish, M., and Donoho, D.~L., \enquote{The Optimal Hard Threshold for Singular
  Values is $4/\sqrt {3}$,} \emph{IEEE Transactions on Information Theory},
  Vol.~60, No.~8, 2014, pp. 5040--5053.
\newblock \doi{10.1109/TIT.2014.2323359}.

\bibitem[{Hemati et~al.(2017)Hemati, Rowley, Deem, and Cattafesta}]{hemati2017}
Hemati, M.~S., Rowley, C.~W., Deem, E.~A., and Cattafesta, L.~N.,
  \enquote{De-biasing the dynamic mode decomposition for applied Koopman
  spectral analysis of noisy datasets,} \emph{Theoretical and Computational
  Fluid Dynamics}, Vol.~31, No.~4, 2017, pp. 349--368.
\newblock \doi{10.1007/s00162-017-0432-2}.

\bibitem[{Scherl et~al.(2020)Scherl, Strom, Shang, Williams, Polagye, and
  Brunton}]{scherl2020}
Scherl, I., Strom, B., Shang, J.~K., Williams, O., Polagye, B.~L., and Brunton,
  S.~L., \enquote{Robust principal component analysis for modal decomposition
  of corrupt fluid flows,} \emph{Phys. Rev. Fluids}, Vol.~5, 2020, p. 054401.
\newblock \doi{10.1103/PhysRevFluids.5.054401}.

\bibitem[{Baddoo et~al.(2021)Baddoo, Herrmann, McKeon, Kutz, and
  Brunton}]{baddoo2021}
Baddoo, P.~J., Herrmann, B., McKeon, B.~J., Kutz, J.~N., and Brunton, S.~L.,
  \enquote{Physics-informed dynamic mode decomposition (piDMD),} , 2021.
\newblock \doi{10.48550/ARXIV.2112.04307}.

\bibitem[{Askham and Kutz(2018)}]{askham2018}
Askham, T., and Kutz, J.~N., \enquote{Variable Projection Methods for an
  Optimized Dynamic Mode Decomposition,} \emph{SIAM Journal on Applied
  Dynamical Systems}, Vol.~17, No.~1, 2018, pp. 380--416.
\newblock \doi{10.1137/M1124176}.

\bibitem[{Rowley et~al.(2004)Rowley, Colonius, and Murray}]{rowley2004}
Rowley, C.~W., Colonius, T., and Murray, R.~M., \enquote{Model reduction for
  compressible flows using POD and Galerkin projection,} \emph{Physica D:
  Nonlinear Phenomena}, Vol. 189, No.~1, 2004, pp. 115--129.
\newblock \doi{10.1016/j.physd.2003.03.001}.

\bibitem[{Brunton et~al.(2016)Brunton, Johnson, Ojemann, and
  Kutz}]{brunton2016}
Brunton, B.~W., Johnson, L.~A., Ojemann, J.~G., and Kutz, J.~N.,
  \enquote{Extracting spatial–temporal coherent patterns in large-scale
  neural recordings using dynamic mode decomposition,} \emph{Journal of
  Neuroscience Methods}, Vol. 258, 2016, pp. 1--15.
\newblock \doi{10.1016/j.jneumeth.2015.10.010}.

\bibitem[{Le~Clainche and Vega(2017)}]{clainche2017}
Le~Clainche, S., and Vega, J.~M., \enquote{Higher Order Dynamic Mode
  Decomposition,} \emph{SIAM Journal on Applied Dynamical Systems}, Vol.~16,
  No.~2, 2017, pp. 882--925.
\newblock \doi{10.1137/15M1054924}.

\bibitem[{Kaptanoglu et~al.(2020)Kaptanoglu, Morgan, Hansen, and
  Brunton}]{kaptanoglu2020}
Kaptanoglu, A.~A., Morgan, K.~D., Hansen, C.~J., and Brunton, S.~L.,
  \enquote{Characterizing magnetized plasmas with dynamic mode decomposition,}
  \emph{Physics of Plasmas}, Vol.~27, No.~3, 2020, p. 032108.
\newblock \doi{10.1063/1.5138932}.

\bibitem[{Lutz et~al.(0)Lutz, Kleinert, Waldmann, Koop, Yorita, Dietz, and
  Schulz}]{lutz2022}
Lutz, T., Kleinert, J., Waldmann, A., Koop, L., Yorita, D., Dietz, G., and
  Schulz, M., \enquote{Research Initiative for Numerical and Experimental
  Studies on High-Speed Stall of Civil Aircraft,} \emph{Journal of Aircraft},
  Vol.~0, No.~0, 0, pp. 1--14.
\newblock \doi{10.2514/1.C036829}.

\bibitem[{Kleinert et~al.(2023{\natexlab{a}})Kleinert, Ehrle, Waldmann, and
  Lutz}]{kleinert2023}
Kleinert, J., Ehrle, M., Waldmann, A., and Lutz, T., \enquote{Wake Tail Plane
  Interactions for a Tandem Wing Configuration in High-Speed Stall Conditions,}
  , 2023{\natexlab{a}}.
\newblock \doi{10.48550/ARXIV.2301.05760}.

\bibitem[{Schwamborn et~al.(2008)Schwamborn, Gardner, von Geyr, Krumbein,
  L{\"u}deke, and St{\"u}rmer}]{dlr55519}
Schwamborn, D., Gardner, A., von Geyr, H., Krumbein, A., L{\"u}deke, H., and
  St{\"u}rmer, A., \enquote{Development of the TAU-Code for aerospace
  applications,} \emph{50th NAL International Conference on Aerospace Science
  and Technology}, 2008.
\newblock \urlprefix\url{https://elib.dlr.de/55519/}.

\bibitem[{Ehrle et~al.(2020{\natexlab{a}})Ehrle, Waldmann, Lutz, and
  Kr{\"a}mer}]{ehrle2020}
Ehrle, M., Waldmann, A., Lutz, T., and Kr{\"a}mer, E., \enquote{Simulation of
  transonic buffet with an automated zonal DES approach,} \emph{CEAS
  Aeronautical Journal}, Vol.~11, No.~4, 2020{\natexlab{a}}, pp. 1025--1036.
\newblock \doi{10.1007/s13272-020-00466-7}.

\bibitem[{Ehrle et~al.(2020{\natexlab{b}})Ehrle, Waldmann, Lutz, and
  Kr{\"a}mer}]{ehrle2020b}
Ehrle, M.~C., Waldmann, A., Lutz, T., and Kr{\"a}mer, E., \enquote{An Automated
  Zonal Detached Eddy Simulation Method for Transonic Buffet,} \emph{Progress
  in Hybrid RANS-LES Modelling}, edited by Y.~Hoarau, S.-H. Peng,
  D.~Schwamborn, A.~Revell, and C.~Mockett, Springer International Publishing,
  Cham, 2020{\natexlab{b}}, pp. 271--281.

\bibitem[{Kleinert et~al.(2023{\natexlab{b}})Kleinert, Stober, and
  Lutz}]{Kleinert2023a}
Kleinert, J., Stober, J., and Lutz, T., \enquote{Numerical simulation of wake
  interactions on a tandem wing configuration in high-speed stall conditions,}
  \emph{CEAS Aeronautical Journal}, Vol.~14, No.~1, 2023{\natexlab{b}}, pp.
  171--186.
\newblock \doi{10.1007/s13272-022-00634-x}.

\bibitem[{Weiner and Semaan(2021)}]{weiner2021}
Weiner, A., and Semaan, R., \enquote{flowTorch - a Python library for analysis
  and reduced-order modeling of fluid flows,} \emph{Journal of Open Source
  Software}, Vol.~6, No.~68, 2021, p. 3860.
\newblock \doi{10.21105/joss.03860}.

\bibitem[{Sashidhar and Kutz(2022)}]{sashidhar2022}
Sashidhar, D., and Kutz, J.~N., \enquote{Bagging, optimized dynamic mode
  decomposition for robust, stable forecasting with spatial and temporal
  uncertainty quantification,} \emph{Philosophical Transactions of the Royal
  Society A: Mathematical, Physical and Engineering Sciences}, Vol. 380, No.
  2229, 2022, p. 20210199.
\newblock \doi{10.1098/rsta.2021.0199}.

\bibitem[{Dylewsky et~al.(2019)Dylewsky, Tao, and Kutz}]{dylewsky2019}
Dylewsky, D., Tao, M., and Kutz, J.~N., \enquote{Dynamic mode decomposition for
  multiscale nonlinear physics,} \emph{Phys. Rev. E}, Vol.~99, 2019, p. 063311.
\newblock \doi{10.1103/PhysRevE.99.063311}.

\end{thebibliography}
